\documentclass{amsart}
\usepackage{amsmath,amsfonts,amscd,amssymb,epsfig,euscript,bm,enumerate}
\usepackage{amsxtra,mathrsfs,xcolor,titletoc} 
\usepackage{mathrsfs}
\usepackage{tikz}
\usepackage{pgf,tikz-cd}	
\usetikzlibrary{decorations.markings}
\usepackage{caption,slashed}
\numberwithin{equation}{section}
\usepackage{pdfsync}
\usepackage{bm}
\usepackage{calligra}
\usepackage[T1]{fontenc}
\newtheorem*{theorem*}{Theorem}

\theoremstyle{definition}
{}

\theoremstyle{remark} 
\newtheorem{remark}{Remark}

\newcommand{\field}[1]{\ensuremath{\mathbb{#1}}}
\newcommand{\CC}{\field{C}}

\newcommand{\HH}{\field{H}}

\newcommand{\RR}{\field{R}}

\newcommand{\ZZ}{\field{Z}}
\DeclareMathOperator{\Tr}{\mathrm{Tr}}

\newcommand{\beq}{\begin{equation}\begin{aligned}}
\newcommand{\eeq}{\end{aligned}\end{equation}}
\newcommand{\eqdef}{\overset{\text{def}}{=}}

\newcommand{\al}{\alpha}

\newcommand{\be}{\beta}
\newcommand{\del}{\delta}
\newcommand{\g}{\gamma}

\newcommand{\bk}{\backslash}
\newcommand{\Ga}{\Gamma}

\newcommand{\pa}{\partial}
\newcommand{\la}{\langle}
\newcommand{\ra}{\rangle}

\newcommand{\ov}{\over}
\newcommand{\ep}{\epsilon}

\newcommand{\z}{\bar{z}}
\newcommand{\curly}[1]{\mathscr{#1}}

\newcommand{\cD}{\curly{D}}

\newcommand{\cH}{\curly{H}}

\newcommand{\cX}{\curly{X}}

\newcommand{\lam}{\lambda}

\newcommand{\ga}{\gamma}

\newcommand{\ex}[1]{\langle #1 \rangle}

\newcommand{\Ad}[1]{\text{Ad}_{ #1 }}
\DeclareMathAlphabet{\mathcalligra}{T1}{calligra}{m}{n}
\DeclareFontShape{T1}{calligra}{m}{n}{<->s*[2.2]callig15}{}
\newcommand{\sr}{\mathcalligra{r}\,}

\makeatletter
\@namedef{subjclassname@2020}{\textup{2020} Mathematics Subject Classification}
\makeatother

\usepackage{hyperref}
\hypersetup{colorlinks,citecolor=blue,plainpages=false,hypertexnames=false}

\begin{document}
\title{Supersymmetry and trace formulas II. Selberg trace formula}
\author{Changha Choi}
\address{Perimeter Institute for Theoretical Physics,
Waterloo, Ontario, N2L 2Y5, Canada}
\author{ Leon A. Takhtajan}
\address{Department of Mathematics,
Stony Brook University, Stony Brook, NY 11794 USA; 
\newline
Euler International Mathematical Institute, Pesochnaya Nab. 10, Saint Petersburg 197022 Russia}
\begin{abstract}
By extending the new supersymmetric localization principle introduced in \cite{Choi:2021yuz}, we present a path integral derivation of the Selberg trace formula on arbitrary compact Riemann surfaces, including the case of vector-valued automorphic forms of arbitrary half-integer weight corresponding to Maass Laplacian. We also generalize the method to formulate the Selberg trace formula on generic compact locally symmetric space. 
\end{abstract}
\subjclass[2020]{}
\maketitle
\tableofcontents

\section{Introduction}

Selberg trace formula \cite{selberg1956harmonic} exhibits deep and highly non-trivial spectral/geometric duality on Riemann surfaces of constant negative curvature, which relates the spectrum of Laplace operator and closed geodesics. The information on the eigenvalues of the Laplacian is only accessible by numerical methods for generic Riemann surfaces, and the Selberg trace formula is the only available analytic tool to uncover the detailed structure of the spectrum.

The possible implication in physics is extremely tantalizing, since the trace formula can be reformulated as a non-trivial identity regarding the thermal partition function of the purely bosonic system. The Selberg trace formula can be thought of as a chaotic version of the trace formula as opposed to the Eskin trace formula on compact Lie groups, since the particle moving on compact Riemann surfaces is chaotic both classically and quantum mechanically. Note that most physical systems of utmost interest (from Yang-Mills theory to black holes) are not integrable nor having exactly solvable spectrum. Therefore, it is desirable to understand the precise physical origin of the Selberg trace formula from the quantum mechanical point of view, which might give some salient physical lessons. 

It is important to highlight the remarkable observation made by Gutzwil-ler \cite{gutzwiller1985geometry}, who discovered that the part of the Selberg trace formula related to the hyperbolic elements coincides with the exact semiclassical contribution to the partition function by closed geodesics, up to DeWitt term. This led him to formulate the `Gutzwiller trace formula', which expresses, at the semiclassical level,  the spectrum of a quantum system in terms of periodic orbits of the corresponding classical chaotic system. However, understanding  in this approach the appearance of equally important remaining terms in the Selberg trace formula  was a mystery.

The goal of the present paper is to provide a physical origin of the Selberg trace formula in the light of the path integral formalism. We will demonstrate that the Selberg trace formula, in its precise form, can be derived from the new supersymmetric localization principle, formulated in \cite{Choi:2021yuz} and successfully applied to the Eskin trace formula for compact Lie groups. 

The organization of the paper is the following. In Section \ref{sec:extended}, we streamline and extend the new supersymmetric localization principle presented in \cite{Choi:2021yuz}. In Section \ref{sec:non-compact}, we study the propagator on non-compact semisimple Lie group by applying the new localization principle to certain supersymmetric sigma model, which will be the stepping stone for the rest of the paper. In Section \ref{sec:Selberg}, we carefully examine the relevant model and the corresponding observables that give the Selberg trace formula on compact Riemann surface, and successfully derive it using the localization principle. Finally, in Section \ref{sec:general} we use localization principle to formulate the Selberg trace formula on general compact locally symmetric space, with a non-trivial check for the case of compact hyperbolic 3-manifolds. 

\subsubsection*{Acknowledgments} The first author (C.C.) thanks J. Gomis, Z. Komargodski, K. Lee, N. Lee, and  P. Yi for discussions. The research of C.C. was supported by the Perimeter Institute for Theoretical Physics. Research at Perimeter Institute is supported in part by the Government of Canada through the Department of Innovation, Science and Economic Development and by the Province of Ontario through the Ministry of Colleges and Universities. The second author (L.T.) thanks A. Alekseev, S. Shatashvili, and D. Sullivan for 
discussions.

\section{Extended new supersymmetric localization principle} \label{sec:extended}

Here we formulate a generalized version of the new supersymmetric localization principle, first presented in \cite{Choi:2021yuz}, which is applicable to a large class of non-supersymmetric correlation functions of supersymmetric QFT in any spacetime dimension.

Specifically, consider a supersymmetric QFT which has at least one Hermitian supercharge $\hat Q$. We are interested in general non-supersymmetric correlation function made out of operators (or defects) with arbitrary codimensions. This situation is quite natural, since usual supersymmetric correlation functions with $(-1)^F$ can be thought of as insertions of  codimension 1 (topological) defects.

Let 
\beq \label{eq:genpi-0}
\ex{\hat{ \mathcal O}(\{\mathcal M_0,\mathcal M_1,\dots, \mathcal M_d\})	}_E=\bm\int  \, { \mathcal O}(\{\mathcal M_0,\mathcal M_1,\dots, \mathcal M_d\}) e^{-S_E}\curly D \mu,
\eeq
be generic correlation function of certain $d$-dimensional reflection positive Euclidean QFT, 
where $M_j$ denotes a union of the domain of all codimension $d-j$ operators and $\mathscr D\mu$ is some `integration measure'. 
We assume that the theory is supersymmetric, $\delta S_{E}=0$, where $\delta$ is the supersymmetry generator, which in the operator formalism is encoded into the definition of the bracket, and  $\delta(\mathscr D\mu)=0$.

A correlation function \eqref{eq:genpi-0}  is called supersymmetric if the integrand  is $\delta$-closed, 
$$\delta { \mathcal O}(\{\mathcal M_0,\mathcal M_1,\dots, \mathcal M_d\})=0.$$

The standard localization principle is the statement that supersymmetric correlation function is invariant under the change $S_E \mapsto S_E+\lambda\del V$ with an arbitrary positive parameter $\lambda$, where a deformation $V$ is such $\del^2 V=0$ and  $\del V$ satisfies the standard positivity property. 

The main idea of \cite{Choi:2021yuz} was to extend this deformation invariance to the case of non-supersymmetric observables and to formulate a new supersymmetric localization principle. This was achieved for systems having fermionic zero modes in the action, and to have a non-vanishing correlation function, we must require that $\mathcal O$ saturates all fermionic zero modes. We formulate it as a statement that non-supersymmetric path integral \eqref{eq:genpi-0} is invariant under the change $S_E\rightarrow S_E+\lambda \del V$ where, in addition to  $\del^2V=0$, the deformation $V$ has the property that both $V \delta \mathcal O$ and $\del V$ do not saturate fermionic zero modes.
It is easy to establish this generalized localization principle along the lines in \cite[Section 2]{Choi:2021yuz}.

It is important to observe that the localization argument outlined above works not only for the Euclidean path integral, but is also applicable to the Lorentzian path integral, where the weighting function is $e^{iS_L}$ with Minkowski (real time) action $S_{L}$. This framework is useful when the observable $\mathcal{O}$ doesn't make sense in Euclidean time, or even if it does, there is no natural Euclidean path integral associated with it (without using some analytic continuation). Here we consider an example of the Lorentzian path integral
\beq \label{eq:genpi}
\ex{\hat{ \mathcal O}(\{\mathcal M_0,\mathcal M_1,\dots, \mathcal M_d\})	}_L=\bm\int  { \mathcal O}(\{\mathcal M_0,\mathcal M_1,\dots, \mathcal M_d\}) e^{iS_L}\,\curly D\mu,
\eeq
which is invariant under the change $S_L\rightarrow S_L +\lambda \del V$, provided the deformation $V$ satisfies the conditions stated above.

We remark that this  approach is quite natural from both physics and mathematics perspectives. Namely, the path integral was first formulated by Feynman in terms of the physical time, where an integrand has an oscillatory nature. This path integral, in semi-classical approximation, can be evaluated by the stationary phase method, based on the mathematical idea that rapid oscillations of an integrand sum up to a negligible contribution.\footnote{In principle, there could also be a contribution from `complex' critical points which needs the method of steepest descent whose path integral analogue has been discussed in \cite{Witten:2010cx}. Following the standard logic, for all examples considered in the paper it can be argued that only `real' critical points contribute for our choice of the `real' contour, since $\del V=0$ at all possible critical points. It would be interesting to explore cases where such analytic continuation is necessary.  } In mathematics, it  was first observed by Duistermaat and Heckman \cite{duistermaat1982variation} that for a certain class of finite-dimensional oscillatory integrals the stationary phase method  gives an exact answer. This idea was subsequently generalized by Atiyah and Bott \cite{atiyah1984moment} and by Berline and Vergne \cite{berline1983zeros}, and led to a mathematical notion of a localization.

Presented extension of the localization principle to non-supersymmetric observables is quite encouraging. We believe it can be successfully applied to even larger class of non-supersymmetric observables, and it may be based on different conditions than the ones used in this paper. The main lesson is that one can still find more surprises in the supersymmetric path integral. 

 \section{Propagator on non-compact semisimple $G$} \label{sec:non-compact}
 
The main focus of this section is on quantum system describing a bosonic particle moving on a group manifold of a non-compact real semisimple Lie group $G$. Throughout this paper, we endow $G$ by a bi-invariant pseudo-Riemannian metric, determined by\footnote{Note that in the case of a compact Lie group, considered in \cite{Choi:2021yuz}, a bi-invariant Riemannian metric is determined by the negative of the Cartan-Killing form.} the Cartan-Killing form $B$ on the Lie algebra $\frak{g}$ of $G$,

\beq \label{eq:CKmetric}
\ex{u,v}_g\equiv B((L_{g^{-1}})_{*}u, (L_{g^{-1}})_{*}v) \quad\text{for all}\quad u,v\in T_gG,
\eeq
where $L_{g}$ are the left translations on $G$. Note that this pseudo-Riemannian metric $G$ induces left-invariant Riemannian metric on the homogeneous space $G/K$, where $K$ is a maximal compact subgroup of $G$. Indeed, restriction of $B$ to the Lie algebra $\frak{k}$ of $K$ is negative-definite, while its restriction to the orthogonal complement $\frak{p}$ of $\frak{k}$ in $\frak{g}$ (with respect to $B$) 
is positive-definite. We denote $ n=\dim G$ and $d_{\frak{k}}=\dim\frak{k}$, $d_{\frak{p}}=\dim\frak{p}$, so $n=d_{\frak{p}}+d_{\frak{k}}$.

Consider the Hilbert space $\cH_{B}=L^{2}(G, dg)$, with respect to the Haar measure $dg$ on $G$. The Lie algebra $\frak{g}$ acts on functions on $G$ by the left-invariant vector fields, 
\beq
(\hat{u}f)(g)=\left.\frac{d}{dt}\right|_{t=0}f(ge^{tu})\quad\text{for all}\quad  u\in\frak{g},
\eeq
and the universal enveloping algebra $U\frak{g}$ of $\frak{g}$ acts on $\cH_{B}$ by the left-invariant differential operators.
The Laplace operator $\Delta$ of $G$ is a negative of a quadratic Casimir element $C_{2}$ in $U\frak{g}$. Explicitly (see Appendix \ref{A} for the case $G=\mathrm{SL}(2,\RR)$), 
\beq
\Delta=-\sum_{a=1}^{d_{\frak{p}}}\hat{e}^{2}_{a}\;\; +\sum_{a=d_{\frak{p}}+1}^{n}\hat{e}^{2}_{a},
\eeq
where according to the decomposition $\frak{g}=\frak{p}\oplus\frak{k}$,  an orthogonal basis $e_{a}$ of $\frak{g}$ satisfies $B(e_{a},e_{a})=1$ for $a=1,\dots, d_{\frak{p}}$ and $B(e_{a},e_{a})=-1$ for $a=d_{\frak{p}}+1,\dots,n$. 

The corresponding real time quantum Hamiltonian $\hat{H}=\Delta/2$ is an unbounded self-adjoint operator on $\cH_{B}$ having an absolutely continuous spectrum of infinite multiplicity filling the real line $\RR$.
The evolution operator $U(iT)=e^{-i\hat{H}T}$ makes a perfect sense as unitary operator with a distributional kernel, while in Euclidean time the operator $e^{-\hat{H}T}$ still exists, but is an unbounded self-adjoint operator on $\cH_{B}$.
It was shown by Krausz and Marinov \cite{Krausz:1997gw} that the real time propagator, a distributional kernel of $U(iT)$, exhibits a similar `almost` semi-classical exactness as in the case of compact Lie groups.

Here we demonstrate that the propagator of a quantum mechanical bosonic particle on a non-compact semisimple Lie group $G$ can be understood by the localization principle, exactly as in the case of compact Lie group \cite{Choi:2021yuz}. The setup is similar to the compact case; the main difference is that because Cartan-Killing metric is indefinite, we need to  work with the real time and not with the Euclidean time as in  \cite{Choi:2021yuz}. Relatedly, we cannot naively use the fermionic Lagrangian ${i\ov 2}\ex{\psi, \dot \psi}$ with  $\psi=\psi^{a}T_{a}\in\Pi L\frak{g}$ in the 
supersymmetric action of \cite{Choi:2021yuz}. The reason is quite simple: canonical (graded) commutation relations $[\hat \psi^a,\hat \psi^b ]=g^{ab}$ can not be represented in a positive-norm Hilbert space, because the metric on the subspace $\mathfrak k$ of the real Lie algebra $\frak{g}$ is negative-definite.

However, this can be easily overcome by setting $(\hat \psi^a)^\dagger=\hat \psi^a$ for $a=1,\dots, d_{\mathfrak p}$ and $(\hat \psi^a)^\dagger=-\hat \psi^a$ along $a=d_{\mathfrak p}+1,\dots, n$. On the path integral side, this corresponds to a choice of integration contour for fermions such that $\psi=\psi^aT_a\in \Pi L\mathfrak p \oplus   i \Pi L\mathfrak k$.

After straightening out this subtlety, we are allowed to use the supersymmetric action on $G$ generalizing the compact case \cite{Choi:2021yuz} which is
 \beq \label{eq:lagG}
 \mathcal L={1\ov 2}\ex{J,J}+{i\ov 2}\ex{\psi,  \dot \psi},
 \eeq
 where $J=g^{-1}\dot{g}\in L\frak{g}$ and it has a supersymmetry
 \beq \label{eq:susyG}
 ~&\del g=ig\psi,
 \\&\del \psi=-J-i\psi\psi.
 \eeq
Exactly as in Proposition 2 in \cite{Choi:2021yuz},  the quantum supercharge $\hat{Q}$ and the Hamiltonian operator $\hat{H}$ satisfy the SUSY algebra and are given by
\beq\label{eq:Q-H-G}
~& \hat{Q}=\ex{\hat{L},\hat\psi}+{i\ov 6}\ex{\hat\psi,[\hat\psi,\hat\psi]}
\\&\hat H=\hat Q^2={1\ov 2}\Delta+{1\ov 48}f_{abc}f^{abc}\hat I={1\ov 2}\Delta+{R \ov 12}\hat I, 
\eeq
where $\hat L\equiv \hat l_a T^a$ (see \cite{Choi:2021yuz}) and $R=-n/4$ is a scalar curvature of the pseudo-Riemannian metric on $G$.

As in \cite{Krausz:1997gw}, we are interested in the propagator between two points connected by the exponential map, the identity element $1_{G}$ of $G$ and the element  $e^{h}$, where $h\in \mathfrak g$. We denote $\mathfrak h\subset \mathfrak g$ to be a Cartan subalgebra containing $h$.

 In compact Lie group case, the propagator was introduced  by the formula (4.25)  in \cite{Choi:2021yuz}, where both bosonic and fermionic traces were involved, which results in the integration over $\Pi TLG$, where $LG$ is the free loop space of $G$. However, in the  non-compact case the volume of $G$ is infinite, so we need to consider the following representation for the propagator
\beq \label{eq:piG}
~& e^{{i \ex{\rho,\rho} T\ov 2}} \langle e^h| e^{-{i\Delta T\ov 2}}|1_G \rangle
\\& =\text{Tr}_{\curly H_F}\left( (-1)^F c_n \hat \chi_1\dots \hat \chi_{n}  \langle 1_{G}  | e^{-i\hat Ht+i\ex{h,\hat r}} |1_{G}\rangle \right)
\\&=\bm\int \limits_{\Pi T \Omega G} e^{iS^h} \cD g \cD\psi,
\eeq
where $\hat \chi^a$ are fermionic zero modes associated with $\psi^a$, normalized such that $[\hat \chi^a,\hat \chi^b]=\del^{ab}$, i.e. $\hat\chi^a=\hat \psi^a$ for $a=1,\dots, d_{\mathfrak p}$ and $\hat\chi^a=i\hat \psi^a$ for $a= d_{\mathfrak p}+1,\dots,n$, $\Omega G$ is the space of based loops on $G$, $c_n=\pm i^{n(n-1)/2}$ is an overall phase to ensure the trace identity (see discussions in \cite{Choi:2021yuz}), and $S^{h}$ is the Minkowski action
\beq
 S^h={1\ov 2}\int_0^T (\ex{J^h,J^h}+i \ex{\psi,\dot \psi}) dt,\quad J^h= J+{1\ov T}\text{Ad}_{g^{-1}}h.
\eeq
 Note that the integration in \eqref{eq:piG} goes over $PT\Omega G$, so the product $\chi_{1}\cdots\chi_{n}$ of  fermionic zero modes drop out of the path integral. As  in \cite{Choi:2021yuz}, in the derivation we used the Freudenthal-de Vries's `strange formula' $n=24\ex{\rho,\rho}$,\footnote{We still denote the naturally induced Cartan-Killing form on the dual space $\mathfrak h^{*}_{\mathbb C}$ as $\ex{~,~}$.} where $\rho\in \mathfrak h_{\mathbb C}^*$ is a Weyl vector (a half-sum of positive roots of $\mathfrak g_{\mathbb C}$).
 
 \begin{remark}\label{measure} The bosonic and fermionic `integration measures' are 
$$\cD g=\prod_{0\leq t\leq T}dg(t)\quad\text{and}\quad \cD\psi=\prod_{0\leq t\leq T}d\psi(t),$$
where 
$$dg=\sqrt{|\det g(x)|}\,dx^{1}\cdots dx^{n}$$ 
is the volume form of the pseudo-Riemannian metric on $G$ associated with the Cartan-Killing form (see formula \eqref{Haar} in Appendix \ref{A}) and 
$$d\psi=\frac{1}{\sqrt{|\det g(e)|}}d\psi^{1}\cdots d\psi^{n},\quad \psi=\psi^{a}T_{a}\in\Pi\frak{g}.$$
\end{remark}

It should be noted that original supersymmetry $\del$ on $\Pi TLG$, given by the formulas \eqref{eq:susyG}, does not preserve the domain of integration $\Pi T\Omega G$, so it cannot be used for the localization of the path integral \eqref{eq:piG}. The space of based loops $\Omega G$ can be thought of as an infinite-dimensional symplectic manifold, so finding a  $\Pi T\Omega G$ preserving supersymmetry transformation is equivalent to finding a Hamiltonian vector field generating the circle action. Intuitively it is not obvious that there is such transformation that preserves the boundary condition $g(0)=1_{G}$. However, we found that the action $S^h$ is invariant with respect to following deformed supersymmetry, which is compatible with this constraint, 
\beq \label{eq:dsusyG}
~&\del_{0} g=ig \psi,
\\& \del_{0} \psi=-J+J(0) -i\psi\psi.
\eeq
Note that $\del_{0}$ is completely `local'  in the sense that it treats the pair $(g,\psi)$ as an element of $\Pi T \Omega G$. Thus we may apply the standard equivariant localization principle with the supersymmetry $\del_{0}$ and to choose the following deformation\footnote{Note that the fixed points of \eqref{eq:dsusyG} are closed geodesics, so one can use the Duistermaat-Heckman's exact stationary phase approximation directly. For compact $G$, this approach was envisioned in \cite{atiyah1985circular} and was based on the earlier work \cite{atiyah1983convexity}; later it was explicitly worked out in \cite{Picken:1988ev} to give the Eskin trace formula,  up to the DeWitt term (see \cite{Choi:2021yuz} for the further discussion).}
\beq
~&V=-{1\ov 2}\int _0^T \ex{\dot J^h,\dot \psi}dt, 
\\& \del_{0} V=\int _0^T \left({1\ov 2}\ex{\dot J^h,\dot J^h} +{i\ov 2} \ex{\dot \psi, (\pa_t+\text{ad}_{J^h})\dot \psi}\right)dt,
\eeq
which satisfying $\del_{0}^2 V=0$.

The important difference in this case is that the path integral is oscillating and the domain of integration is the based loop space on a `non-compact' target. Hence we necessarily have to check the potential subtlety arising from the boundary in the paths space, which might break the invariance of the path integral for invariant deformation. To verify that it is not the case, we use the identity
\beq
\bm\int \limits_{\Pi T \Omega G}  e^{i(S^h+\lambda \del V)} \cD \mu =\bm\int  \limits_{\Pi T \Omega G}  e^{iS^h }\cD \mu +\int_0^\lambda d \tilde \lambda \bm\int  \del \left(  i V e^{i (S+\tilde \lambda \del V)}\right)\cD \mu,
\eeq
where $\cD\mu=\cD g\cD\psi$. 
It follows from the Stokes theorem that the last term is a contribution from the boundary of the space of fields. Since in our case there is no bosonic zero mode, this boundary contribution vanishes due to the highly oscillating nature of the integrand, like in the classical Riemann-Lebesgue lemma.
As a result, the path integral is invariant under the transformation $S^h\rightarrow S^h+\lambda \del V$ with arbitrary `real' $\lambda $. 

Therefore, as we take the limit $\lambda \rightarrow \infty$, the path integral, because of the highly oscillatory nature of the deformation, localizes onto the stationary points $\dot J^h=0$. The computation is completely analogous to that in \cite{Choi:2021yuz}, and for a regular, semisimple $h\in\frak{g}$ we obtain\footnote{
We remark that this approach, based on using the deformed supersymmetry in $\Pi T \Omega G$, is also equally applicable to the case when $G$ is a compact Lie group. The approach taken in \cite{Choi:2021yuz} was to employ localization at the level of $\Pi T L G$, which uses the new localization principle.}
\begin{equation}\label{prop-on-G}
\begin{gathered}
\langle e^h| e^{-{i\Delta T\ov 2}}|1_G \rangle \\={ e^{-{i  \ex{\rho,\rho} T \ov 2}}  \ov (2\pi i T)^{d_{\mathfrak p}/2} (-2\pi i T)^{d_{\mathfrak k}/2}}\sum \limits_{\gamma \in \Gamma^h} \prod_{\al\in  R^{+}_{\frak{g}_{\mathbb C}, \mathfrak h_{\mathbb C}}  }  {{1\ov 2}\ex{\al, h+\gamma }\ov \sinh{{1\ov 2}\ex{\al,h+\gamma }}}e^{\frac{i\ex{h+\gamma,h+\gamma}}{2T}}.
\end{gathered}
\end{equation}
Here $\Gamma^h$ is the characteristic lattice in the Cartan subalgebra $\mathfrak h $ containing $h$ and $R^{+}_{\frak{g}_{\mathbb C} , \mathfrak h_{\mathbb C} } $  is a set of positive roots in $\frak{g}_{\mathbb C}$ with respect to $\mathfrak h_{\mathbb C}$.\footnote{
Throughout the paper, we adopt the standard conventions on $\mathfrak g$ and $G$, used in \cite{Choi:2021yuz, Kirillov-book}.} This compactly reproduces results in \cite{Krausz:1997gw} and reinforces the localization procedure described above.\footnote{Note that the overall prefactor is slightly different from \cite{Krausz:1997gw}, where \cite{Krausz:1997gw} has ${(2\pi i T)^{-n/2}}$ in terms of our convention. This is because \cite{Krausz:1997gw} used the measure $dg=\sqrt{\det g(x)}\,dx^{1}\cdots dx^{n}$ without the absolute value of the determinant. 
}

\begin{remark}  \label{remark:dim}  Note that prefactors of $(2\pi i)^{d_{\mathfrak p}/2} (-2\pi i)^{d_{\mathfrak k}/2}$ in the denominator of \eqref{prop-on-G} are Lorentzian (Pseudo-Riemannian) infinite-dimensional analogues of the prefactor $(2\pi )^{\dim M/2}$ in a Gaussian type integral over the bosonic and fermionic modes over a finite-dimensional
 manifold $M$ (cf. \cite[Remark 6]{Choi:2021yuz} for the case of Euclidean path integral). Moreover, according to $\prod_{n=1}^\infty z=1/\sqrt{z},~ -\pi< \text{arg} z\leq \pi $ as in \cite{Choi:2021yuz}, so
 $$\prod_{n=1}^\infty (-1)=e^{-\frac{\pi i}{2}},\quad \prod_{n=1}^\infty i=e^{-\frac{\pi i}{4}}\quad\text{and}\quad \prod_{n=1}^\infty (-i)=e^{\frac{\pi i}{4}}.$$
\end{remark}

  \section{Selberg trace formula on compact Riemann surfaces} \label{sec:Selberg}
  
 The Selberg trace formula relates the spectrum of the Laplace operator of constant negative curvature metric on a Riemann surface with the geodesic length spectrum. By the uniformization theorem, such Riemann surface can be realized as a double coset $X=\Gamma \backslash G /K=\Gamma \backslash \mathbb H$, where $G=\mathrm{SL}(2,\RR)$, $K=\mathrm{SO}(2)$ and $\HH=G/K$ in the Lobachevsky (hyperbolic) plane. Here $\Gamma$ is a cofinite discrete subgroup of $G$, and in this paper we assume that $\Gamma$ is cocompact, so $X$ is a compact Riemann surface (an orbifold Riemann surface, if $\Gamma$ contains elements of finite order).

Here we advocate the point of view that physically it is more advantageous to consider the gauged sigma model on the target $\Gamma\backslash G$ instead of the sigma model on the Riemann surface $X$.  
In particular, this is because $\Gamma \backslash G$ is always smooth, while $X=\Gamma \backslash G /K$ may have orbifold singularities.

\begin{remark} \label{remark1} Note that the group $\Gamma$ acts on $\HH$ by M\"{o}bius transformations, so actually $X\simeq\bar\Gamma\bk\HH$, where $\bar\Gamma=\Gamma/\{\pm I\}\subset\mathrm{PSL}(2,\RR)$ is a Fuchsian group, 
a discrete subgroup of $\mathrm{PSL}(2,\RR)$. This can be understood from the fact that $\{\pm I\}$ is a subgroup of $K=\mathrm{SO}(2)$. Therefore, for any compact orbifold Riemann surface $X$ we can always lift the Fuchsian group $\bar\Gamma$ to a discrete subgroup  $\Gamma$ of $G$ with the property that $-I\in \Gamma$, which will be  assumed in this section.
\end{remark}

\subsection{The gauged sigma model} \label{gauge-sigma}
We start by considering the quotient space $\Gamma\backslash G$, which plays a fundamental role in the representation theory. It is quite natural to ask whether there is a supersymmetric system similar to that in the case of $G$. The answer is remarkably simple, because under the purely bosonic projection $\pi_{\Gamma}:G\rightarrow \Gamma \backslash G$ the Lagrangian \eqref{eq:lagG} remains supersymmetric  with the same supersymmetry \eqref{eq:susyG}. After the projection, the Hilbert space becomes 
$$\curly H_{\Gamma\bk G} = L^2(\Gamma \bk G, dg) \otimes \curly H_{F,\mathfrak g}.$$

Physical realization of the right quotient by $K=\mathrm{SO}(2)$ is achieved by gauging the global symmetry $g\rightarrow gk$, $\psi\rightarrow \text{Ad}_{k^{-1}} \psi$, where $k\in K$, by a connection $A=A^{3}T_{3}$ in the principal $K$-bundle over $S^{1}_{T}$. The simplest such extension that preserves the supersymmetry is given by the gauged sigma model on $\Gamma\backslash G$ with the action
\beq\label{action-L}
 S_{A}[g,\psi]=\int_{0}^{T}\mathcal{L}_{0}\,dt.
 \eeq
 Here 
\beq\label{Lagrangian}
 \mathcal L_0={1\ov 2}\ex{J_A,J_A}+{i\ov 2}\ex{\psi,\pa^{A}_{t} \psi},
 \eeq
 where  
 $$J_A=J-A\quad\text{and}\quad \pa^{A}_{t}=\pa_t+\text{ad}_A$$ 
 are gauge covariant objects. The supersymmetry \eqref{eq:susyG} also gets modified
  \beq \label{eq:susyselberg}
 ~&\del g=ig\psi,
 \\&\del \psi=-J_A-i\psi\psi,
 \\& \del A=0.
 \eeq

However, the physical spectrum of the quantum system with the Lagrangian $\mathcal{L}_{0}$ no longer corresponds to the eigenstates of a purely bosonic Laplacian on a Riemann surface $X$ since
the fermionic part is being gauged as well. To see this, we first perform canonical quantization on $\Gamma \bk G$ and impose the gauge constraint later. In the first step, we can effectively set $A=0$ and hence the quantum supercharge $\hat{Q}$ and the Hamiltonian $\hat{H}$ are given by the same formulas \eqref{eq:Q-H-G}.

Now to impose a gauge invariance, we observe that it follows from \eqref{Lagrangian} that Lagrangian $\mathcal{L}_{0}$ has no kinetic term for $A$, so we have classical Gauss law
\beq\label{Gauss-cl}
C_{0}:~J^{3}_{A}+2i\psi^{1}\psi^{2}=0.
\eeq
Quantum mechanically it is realized as the following constraint on the Hilbert space\footnote{Note that we made a specific choice of a quantization scheme when we promote the classical Gauss law \eqref{Gauss-cl} to the quantum Gauss law \eqref{Gauss}.
%Any other choice of a quantization scheme amounts to a shift $\mathcal{L}_{0} \mapsto\mathcal{L}_{n}$. 
 }
\beq\label{Gauss}
\hat C_0: ~\{\hat{T}_{3}-2\hat N_{F,\mathfrak p}+\hat I=0\},
\eeq
where $\mathfrak p=\mathfrak g/\mathfrak k$ and $\hat N_{F,\mathfrak p}$ is the fermion number operator in the subspace $\curly H_{F,\mathfrak p}$ of the fermion Hilbert space $\curly H_{F,\frak{g}}$. 

Indeed, $\mathcal H_{F,\mathfrak p}=\CC^{2}$ and fermion operators $\hat\psi^{1}$ and $\hat\psi^{2}$ are normalized as follows (see  \cite{Choi:2021yuz}  and Appendix \ref{A})
 \beq\label{f-norm}
 [\hat\psi^{1}, \hat\psi^{1}]=g^{11}=\frac{1}{8} \quad\text{and}\quad [\hat\psi^{2}, \hat\psi^{2}]=g^{22}=\frac{1}{8},
 \eeq
 so in the standard basis of $\CC^{2}$ where $\hat{N}_{F,\frak{p}}=-\frac{1}{2}(\sigma_{3}-I)$ is diagonal, we have
 \beq\label{f-oper}
 \hat\psi^{1}=\frac{1}{4}\sigma_{1},\,\;\hat\psi^{2}=\frac{1}{4}\sigma_{2} \quad\text{and}\quad 2i\hat\psi^{1}\hat\psi^{2}=-\frac{1}{8}\sigma_{3}.
 \eeq\label{N-number}
The current $J^{3}_{A}$ is quantized as in \cite{Choi:2021yuz}, since $g_{33}=\la T_{3}, T_{3}\ra=-8$ we have
$$\hat{J}^{3}_{A}=-\frac{1}{8}(\hat{J}_{A})_{3}=\frac{i}{8}\hat{T}_{3},$$
which gives equation \eqref{Gauss}.

Thus the physical Hilbert space $\curly H_0$ consists of the null space of $\hat C_0$ in the unconstrained Hilbert space $\curly H_{\Gamma\bk G}$.  Representing element in $L^{2}(\Gamma\bk G)\otimes\curly H_{F,\mathfrak p}=L^{2}(\Gamma\bk G)\otimes\CC^{2}$ as a two-component column vector 
$f=\begin{pmatrix}f_{+} \\f_{-}\end{pmatrix}$ and using  that $\hat{T}_{3}=\dfrac{\partial}{\partial\theta}$, we rewrite equation $\hat C_{0}f=0$ as
$$\frac{\partial f}{\partial \theta}=i\sigma_{3}f,$$
so $f_{\pm}(ge^{i\theta T_{3}})=e^{\pm i\theta}f_{\pm}(g)$. This gives a remarkable identification
\beq
~&\curly H_0=(\curly H_{\Gamma\bk G})/\hat{C}_0=L^2_{1/2}(X)\oplus L^2_{-1/2}(X)\oplus \cH_{F,\mathfrak k}.
\eeq
Here $L^2_m(X)$ for $m\in\frac{1}{2}\ZZ$ is the Hilbert space of functions on $\HH$ satisfying
\begin{equation}\label{weight-m}
f\left(\frac{az+b}{cz+d}\right)\frac{|cz+d|^{2m}}{(cz+d)^{2m}}=f(z),\quad \g=\begin{pmatrix} a &b\\ c&d\end{pmatrix}\in\Gamma,
\end{equation}
and square integrable on a fundamental domain of $\Gamma$ in $\HH$ with respect to the hyperbolic area form $y^{-2}dxdy$ (see Appendix \ref{A}), and $\cH_{F,\mathfrak k},$ is the fermionic Hilbert space along $\mathfrak k$.\footnote{Although $d_{\mathfrak k}=1$, we can innocuously extend the definition of the fermionic Hilbert space as discussed in \cite{Choi:2021yuz}. } Restriction of $\Delta$ to the subspaces $L^2_{\pm 1/2}(X)$ coincides with the Maass Laplacians of weight $\pm 1/2$  (see Appendix \ref{A}). In other words, the quantum system corresponding to the Lagrangian $\mathcal{L}_{0}$ describes the Maass Laplacians $D_{\pm\frac{1}{2}}$ of weights $\pm 1/2$ on $X$, together with a single free Majorana fermion along $\mathfrak k$ direction. 

However, the quantum system associated with the Lagrangian $\mathcal{L}_{0}$ is actually anomalous! Namely, since $-I \in \Gamma$, we see that for $m=\pm 1/2$ putting $\gamma=-I$ into \eqref{weight-m}  gives
$f(z)=-f(z)$, so $f(z)=0$ and instead of $L_{\pm 1/2}^2(X)$ we get a zero dimensional space $\{0\}$. In other words, the physical Hilbert space is empty because one can not obey the Gauss law \eqref{Gauss}.

\subsection{Wilson loops and defects}\label{W-defects}

The inconsistency of the Gauss law for  a system with fermions coupled to dynamical gauge fields is a typical  manifestation of an anomaly in the Hamiltonian approach \cite{Faddeev:1984jp,Faddeev:1984ung,Nelson:1984gu, Bardeen:1982wj,Delmastro:2022pfo}. In our case, it can be understood as a `t Hooft anomaly of $\mathbb Z_2$ one-form global symmetry in quantum mechanics.

We recall that a gauge theory in $d$ space-time dimensions with a gauge group $G$ with a center $\mathcal Z(G)$ has an (electric) one-form global symmetry $ \mathcal Z^{(1)} \subset \mathcal Z(G)$ whenever there are no matter fields charged under $\mathcal Z^{(1)}$, and hence the theory can be consistently put on a $G/ \mathcal Z^{(1)}$-bundle \cite{Gaiotto:2014kfa}. On a Wilson loop in a representation charged under $\mathcal Z^{(1)}$, this one-form symmetry acts  by a non-trivial phase.

In our case  the gauge group is $K=\mathrm{SO}(2)$ and the matter fields $g\in \Gamma\bk G$ have charge $2$, while $\psi \in \mathfrak g$ decomposes into charge $2$ and $0$ under $\mathrm{SO}(2)$, so the one-form symmetry is $\mathcal Z^{(1)}= \mathbb Z_2$. This $\mathbb Z_2$ symmetry acts on a Wilson line with the charge $n$ representation of $\mathrm{SO}(2)$ by a phase $(-1)^n$.

Gauging the $\mathbb Z_2$ one-form symmetry changes the structure group to $\mathrm{PSO}(2)=\mathrm{SO}(2)/\mathbb Z_2$. In our case the local form of the Lagrangian $\mathcal{L}_{0}$ and the Gauss law is insensitive to the global structure of the gauge bundle, so there is no gauge-invariant physical state. Moreover, there is no local gauge-invariant counter-terms that makes theory non-anomalous, so the system on the $\mathrm{PSO}(2)$-bundle suffers from a gauge anomaly and is physically inconsistent.

On the other hand, for our original system with the Lagrangian  $\mathcal{L}_{0}$ and the gauge group $K=\mathrm{SO}(2)$, the inconsistency of the Gauss law is related to the `t Hooft anomaly, so we can make the system non-anomalous by adding a gauge-invariant counterterm. It has the form $nA^{3}$, where $n$ is an odd integer. This defines non-anomalous Lagrangians
\beq
\mathcal L_n=\mathcal L_0+nA^3,\quad\text{where}\quad n=2k+1,\quad k\in \mathbb Z.
\eeq
Note that $\mathcal L_n$ is still invariant under the supersymmetry \eqref{eq:susyselberg}. This counter-term can be also viewed as a one-dimensional version of the Chern-Simons term of level $n$. Equivalently, it can be thought of as addition to the action of a temporal Wilson loop of charge $n$,
 $$\mathcal W_{n}(A)=e^{in\int_0^T A^3 dt }.$$

Since $\mathcal W_n(A)$ with an odd integer $n$ has charge $1$ under the $\mathbb Z_2$ one-form symmetry, we see that the path integral with $\mathcal L_0$ suffers from the `t Hooft anomaly for the $\mathbb Z_2$ one-form symmetry which makes the total system $\mathcal L_n$ non-anomalous. The insertion of such defect twists the Hilbert space as
\beq \label{eq:twistHil}
\cH_n=L^2_{{n+1 \ov 2}}(X)\oplus L^2_{{n-1\ov 2}}(X)\oplus \cH_{F,\mathfrak k},
\eeq
which clearly demonstrates that for $n$ being an odd integer the physical Hilbert space $\cH_{n}$ is well-defined and the system is non-anomalous.

We note that for each $n$ the system has a single fermionic zero mode 
$$\psi^3=\frac{1}{T}\int_{0}^{T}\psi^{3}(t)dt,$$
which we will abbreviate as $\psi_{\frak{k}}$, and denote by  $\hat \psi_{\mathfrak k}$ properly normalized fermion along $\mathfrak k$ direction, so that $[\hat \psi_{\mathfrak k},\hat \psi_{\mathfrak k}]=1$ and $\hat \psi_{\mathfrak k}^\dagger =\hat\psi_{\mathfrak k}$.
Therefore, the natural non-supersymmetric observable satisfying extended localization principle discussed in Section \ref{sec:extended} 
is given by
\begin{equation}
\begin{gathered}
I_n\equiv \text{Str}_{\mathcal \cH_n}[\hat\psi_{\mathfrak k}\,e^{-i \hat H_0 T}]
 =\Tr_{L_{{n-1\ov 2}}^2(X)}[e^{-iT \Delta/2}]- \Tr_{L_{{n+1\ov 2}}^2(X)}[e^{-iT \Delta/2}]\\
= {e^{-{i \ex{\rho,\rho}T \ov 2}}\ov\text{vol}(\mathcal G)} \bm{\int} \mathcal W_n(A)\psi_{\mathfrak k}e^{i S_{A}[g,\psi]}\mathscr Dg\mathscr  D\psi \mathscr  DA,
\end{gathered}
\end{equation}
where in the last line, we factor out the volume of the gauge group $\mathcal G$, as it is customary in the gauge theory. 

However, for the purposes of the basic Selberg trace formula, one needs to consider the trace of the operator $e^{-iT\Delta/2}$ in the Hilbert space $L^{2}_{0}(X)$ of square integrable automorphic forms of weight $0$. To achieve this goal, we start with the following extremely naive formula, based on the
`illegal' use of telescoping\footnote{The formula $\sum_{n=1}^{\infty}(a_{n}-a_{n+1})=a_{1}$, which is valid only when $\lim_{n\to\infty}a_{n}=0$.} and geometric series
\begin{equation} \label{eq:mainint}
\begin{gathered}
 \Tr_{L^2_0(X)}[e^{-iT \Delta/2}]\stackrel{?}{=}\sum_{n\in2\mathbb Z+1} {1\ov 2}{\text{sgn}(n)}I_{n}\\ \stackrel{?}{=}{e^{-{i \ex{\rho,\rho}T \ov 2}}\ov\text{vol}(\mathcal G)} \bm{\int}  {1 \ov \mathcal W_{-1}(A)-\mathcal W_1(A)} \psi_{\mathfrak k}e^{i S_{A}[g,\psi]}\mathscr Dg\mathscr  D\psi \mathscr  DA.
 \end{gathered}
\end{equation}
Note that each of  equalities in  \eqref{eq:mainint}  does not really make sense since the series in the first line is not convergent! More serious mathematical objection is that each term $I_n$ doesn't contain any information about the non-zero modes since the spectra of operators $D_{\frac{n+1}{2}}$ and $D_{\frac{n-1}{2}}$ coincide, except for possible finitely many eigenvalues related to zero modes of Hodge Laplace operators (see Appendix \ref{A}).
Physically, this is because $\psi_{\frak{k}}$ is just a free fermion, and hence $I_n$ can be thought of as a standard Witten index for a (twisted) supersymmetric non-linear sigma model on $X$.

However, it is truly amazing that the first and the last term in \eqref{eq:mainint} are equal, and the derivation does not use $I_n$ at all! To demonstrate this, let us start with the last line in \eqref{eq:mainint}, where the domain of integration is
$$L(\Gamma\bk G)\times \Pi L\mathfrak g\times \mathcal{A}.$$
Here $L(\Gamma\bk G)$ and $\Pi L\mathfrak{g}$ are free loop spaces of $\Gamma\bk G$ and $\frak{g}$ (the latter with the reversed parity), and $\mathcal{A}$ is the space of connections on the principle $\mathrm{SO}(2)$-bundle over $S^1_T$.

We fix the gauge in the last line of \eqref{eq:mainint} by setting $A= \bm h /T$ with $\bm h=hT_3$, which for the last term in \eqref{eq:mainint} gives the following expression
\beq \label{eq:derv2}
{e^{-{i \ex{\rho,\rho}T \ov 2}}} \int_{-\pi}^{\pi} {{dh\ov 2\pi}\, } {i \ov 2\sin{ h }} \bm{\int} \psi_{\mathfrak k}e^{iS_f[\bm h](g,\psi)}\mathscr Dg\mathscr  D\psi,
\eeq
where 
\beq \label{eq:Sf}
S_f[\bm{h}](g,\psi)={1\ov 2} \int_0^T \left(\ex{J_{\bm h/T},J_{\bm h/T}}+i\ex{\psi,\pa_t^{\bm h/T} \psi}\right)dt
\eeq
is the fixed gauge action. Here the $2\pi$ in the denominator of \eqref{eq:derv2} comes from the volume of $\mathrm{SO}(2)$. Now the effective path integral for fixed $\bm{h}$ is nothing but a twisted partition function for the supersymmetric particle on $\Gamma\bk G$. Since in \eqref{eq:derv2}--\eqref{eq:Sf}  bosonic and fermionic degree of freedoms are decoupled, we have
\beq \label{eq:derv3}
~&e^{-{i \ex{\rho,\rho}T \ov 2}}\bm{\int} \psi_{\mathfrak k}e^{iS_f[\bm{h}](g,\psi)}\mathscr Dg\mathscr  D\psi 
\\&= \bm\int \psi_{\mathfrak k}e^{-\frac{1}{2}\int_{0}^{T}\ex{\psi,\pa_t^{\bm h/T}\psi} dt}\cD\psi \int_{\Gamma\bk G}   \langle g | e^{-i T \Delta_{\Gamma\bk G}/ 2}| g e^{i \bm h}\rangle dg.
\eeq

In the Hamiltonian representation, the fermion path integral becomes the trace
\beq\label{int=trace}
 \bm\int \psi_{\mathfrak k}e^{-\frac{1}{2}\int_{0}^{T}\ex{\psi,\pa_t^{\bm h/T}\psi} dt}\cD\psi= \text{Tr}_{\cH_{F,\mathfrak g}}\left[(-1)^F \hat\psi_{\mathfrak k}\, e^{-i\hat H_{F}T}\right], 
 \eeq
 where
 \beq\label{H-F-0}
 \hat H_{F}=-\frac{i}{2T}\ex{\hat\psi,\text{ad}_{\bm h} \hat\psi}=\frac{16ih}{T}\hat\psi^{1}\hat\psi^{2}
 \eeq 
 and fermion operators $\hat\psi^{1}$ and $\hat\psi^{2}$ are normalized as in \eqref{f-norm}. Though $d_{\frak{k}}=1$ and Majorana fermions are anomalous, we can use the approach in 
 \cite{Choi:2021yuz}, which effectively reduces the fermion Hilbert space $\cH_{F,\frak{g}}$ to $\CC^{2}$, and the operator $(-1)^{F}\hat\psi_{\frak{k}}$ to $\sigma_{3}$.
 Then using \eqref{f-oper} we obtain
 \beq\label{H-F-2}
 -i\hat{H}_{F}T=-ih\sigma_{3}=-2ih\left(N_{F,\frak{p}}-\frac{1}{2}\right),
 \eeq
so
\begin{equation}\label{Wilson-fermi}
 \text{Tr}_{\cH_{F,\mathfrak g}}\left[(-1)^F \hat\psi_{\mathfrak k}\, e^{-i\hat{H}_{F}T}\right] =-2i\sin h.
\end{equation}
Thus we see that  the fermionic trace \eqref{int=trace} remarkably cancels the factor in \eqref{eq:derv2}, arising from the insertion of $1/(\mathcal W_{-1}(A)-\mathcal W_{1}(A))$ into the path integral \eqref{eq:mainint}.

The final ingredient of the proof is the following basic representation identity 

\beq \label{eq:repid}
\Tr_{L^2_0(X)}\left[e^{-iT \Delta/2}\right]=\int_K {dk\ov \text{vol}(K)}  \int_{\Gamma\bk G }  \langle g | e^{-i T \Delta_{\Gamma\bk G}/2}| g k \rangle dg,
\eeq
 which is valid for any bi-invariant measure $dg$ on $G$, 
and we obtain
\begin{equation} \label{eq:maineq}
\begin{gathered}
Z(iT) =\Tr_{L^2_0(X)}\left[e^{-iT \Delta/2}\right]\\ ={e^{-{i \ex{\rho,\rho}T \ov 2}}\ov\text{vol}(\mathcal G)} \bm{\int}  {1 \ov \mathcal W_{-1}(A)-\mathcal W_1(A)} \psi_{\mathfrak k}e^{i S_{A}[g,\psi]}\mathscr Dg\mathscr  D\psi \mathscr  DA.
\end{gathered}
\end{equation}
Thus we have proved that insertion of a special singular defect into the path integral provides the information about the full spectrum, in contrast with the naive path integral without defects.

Note that  formula \eqref{eq:maineq} should be understood in the sense of generalized functions, since $Z(iT)$ is not a convergent series for real $T$. Instead, we use the boundness of the spectrum of $\Delta$ on a Riemann surface $X$, which makes $Z(iT)$ analytic on the lower half-plane $\text{Im}\,T<0$. This allows us to consider the equality \eqref{eq:maineq} as a boundary limit of analytic functions at the lower half-plane. Therefore, once we evaluate $Z(iT)$ using the localization, it is necessary to analytically continue to the imaginary time in order to obtain the canonical partition function.

\subsection{Localization and the pre-trace formula}\label{pre-trace}
Now we are ready to perform the supersymmetric localization. Since supersymmetry \eqref{eq:susyselberg} doesn't act on the gauge field and commutes with the gauge symmetry, we can first fix the gauge and then apply the supersymmetric localization to the path integral \eqref{eq:derv2}. The bosonic domain of integration in the path integral is a free loop space $L(\Gamma\bk G)$, so we need to describe its connected components.

For this aim, we use the basic fact that for any topological space $\cX$ with a based point $x_{0}$, the based loop space $\Omega_{x_{0}}\cX$ decomposes into the disjoint union of connected components, labeled by the elements of the fundamental group $\pi_{1}(\cX, x_{0})$,
\beq \label{eq:homsequence}
\Omega_{x_{0}}\cX=\bigsqcup_{\xi\in\pi_{1}(\cX,x_{0})}\Omega_{x_{0},\xi}\cX,
\eeq
where  $\Omega_{\xi,x_{0}}\cX$ is the set of all based loops in the same homotopy class $\xi$. In our case $\cX=\Ga\bk G$ and since $p: G\to\Ga\bk G$ is a covering with fibers $\Gamma$, we have an exact sequence of homotopy groups
$$0\to\pi_{1}(G,e)\to\pi_{1}(\Gamma\bk G,\bar{e})\to\pi_{0}(\Ga)\to 0,$$
where $e$ is the unit in $G$ and $\bar{e}$ --- its projection to $\Ga\bk G$. 
The covering $p: G\to\Ga\bk G$ is normal, so the image  $N$ of $\pi_{1}(G,e)$ is a normal subgroup of $\pi_{1}(\Gamma\bk G,\bar{e})$ and 
$$\Ga\simeq\pi_{1}(\Ga\bk G,\bar{e})/N.$$
From here it is easy to derive that 
 \begin{equation}\label{main-decomposition}
L(\Ga\bk G)=\bigsqcup_{g_{0}\in F}\bigsqcup_{[\g]}\left(\bigcup_{\sigma\in\Ga_{\g}\bk\Ga}\mathcal{P}_{\sigma^{-1}\g\sigma}(G,g_{0})\right),
\end{equation}
where $F$ is a connected fundamental domain of $\Ga$ in $G$, $[\g]$ runs over all conjugacy classes in $\Ga$ with representatives $\g$, the group $\Ga_{\g}$ is a centralizer of $\g$ in $\Ga$, and $\mathcal{P}_{\g}(G,g_{0})$ is a space of paths $g(t)$ in $G$ satisfying $g(0)=g_{0}$ and $g(T)=\ga g_{0}$.

Decomposition \eqref{main-decomposition} can be also obtained directly using the basic fact that for any space $\cX$ we trivially have
$$L\cX=\bigsqcup_{x_{0}\in X}\Omega_{x_{0}}\cX.$$
Specifically, in our case $\cX=\Ga\bk G$, and for every $x_{0}\in \cX$ we choose $g_{0}\in G$ such that $p(g_{0})=x_{0}$. Since $\Gamma=p^{-1}(x_{0})$, every other lift of $x_{0}$
is $\sigma g_{0}$, where $\sigma\in\Gamma$. We can always assume that $g_{0}\in F$ for some choice of a connected  fundamental domain of $\Gamma$ in $G$.
By a lifting path property, each $\xi(t)\in\Omega_{x_{0}}\cX$ can be uniquely lifted to a path $g(t)\in\mathcal{P}_{\g}(G,g_{0})$  for some $\g\in\Gamma$. By a homotopy lifting property, loops $\xi_{1},\xi_{2}\in\Omega_{x_{0}}\cX$ are related by a free homotopy with $x_{0}$ moving along $\xi(t)\in\Omega_{x_{0}}\cX$, if $\xi_{1}(t)$ lifts to $g_{1}(t)\in\mathcal{P}_{\g_{1}}(G,g_{0})$, $\xi(t)$ lifts to $g(t)\in\mathcal{P}_{\sigma}(G,g_{0})$ and $\xi_{2}(t)$ necessarily lifts to $g_{2}(t)\in\mathcal{P}_{\g_{2}}(G,\sigma g_{0})$ for some $\g_{1}, g_{2}, \sigma\in\Ga$. Since $\Ga$ is discrete, using  $\mathcal{P}_{\sigma\g\sigma^{-1}}(G, g_{0})=\sigma^{-1}\mathcal{P}_{\g}(G,\sigma g_{0})$ 
we get $\sigma^{-1}\g_{2}\sigma=\g_{1}$, and the decomposition
 \eqref{main-decomposition} follows.

As a result, $Z(iT)$ takes the following form
\beq \label{eq:naturaldecomp}
Z(iT)=\sum_{[\g]} Z_{[\gamma]}(iT),
\eeq
where
\begin{gather*}
Z_{[\g]}(iT) \\={e^{-{i \ex{\rho,\rho}T \ov 2}}} \int_{-\pi}^{\pi} {{dh\ov 2\pi}\, } {i \ov 2\sin{ h }} \sum_{\sigma\in\Ga_{\g}\bk\Ga}\int_{F}d g_{0}\bm\int\limits_{\mathcal{P}_{\sigma^{-1}\g\sigma}(G,g_{0})\times\Pi L\frak{g}}
 \psi_{\mathfrak k}e^{iS_f[\bm h](g,\psi)}\mathscr Dg\mathscr  D\psi
 \end{gather*}
 and $[\g]$ runs over all conjugacy classes in $\Ga$ with representatives $\g$.

The main advantage of \eqref{eq:naturaldecomp} is that the supersymmetry is preserved for each ${[\g]}$, so we can apply the supersymmetric localization for each $Z_{[\g]} (iT)$ separately. Though the path integral \eqref{eq:derv2} is not supersymmetric, the extended localization principle, formulated in Section \ref{sec:extended} is applicable. As in Section \ref{sec:non-compact}, for each ${[\g]}$ we use purely oscillating deformation $V_{[\g]}$; the vanishing of contribution from the boundary of the space of fields is supported by the compactness of $\Gamma \bk G$ and by the arguments in Section \ref{sec:non-compact}.

Thus we arrive at the simplified expression
\beq\label{Z-V}
Z_{[\gamma]} (iT)
&= {e^{-{i \ex{\rho,\rho}T \ov 2}}} \int _{-\pi}^{\pi} {{dh\ov 2\pi}\, } {i \ov 2\sin{ h }} \sum_{\sigma \in\Gamma_\g \bk \Gamma } \int_F \mathcal A_{\sigma^{-1} \gamma \sigma } (g_0;\lambda )dg_0, 
\eeq
where for arbitrary $\ga\in\Gamma$,
\beq\label{A-path-integral}
\mathcal A_{\ga} (g_0;\lambda )=\bm\int \limits _{\mathcal P_\gamma(G;g_0)\times \Pi L\frak{g} } \psi_{\mathfrak k}e^{i(S_f[\bm h](g,\psi)+\lambda \del V_\gamma)}\mathscr Dg\mathscr  D\psi.
\eeq

\begin{remark} \label{2}We observe that if deformation $V_{\g}$ is such that $V_{-\g}=V_{\g}$, than 
$$Z_{[\gamma]}(iT)=Z_{[-\gamma]}(iT),$$
which is intuitively clear because $-I$ is also an element of $K$, therefore $\gamma$ and $-\gamma$ act identically on the homogeneous space $G/K$, which we have physically realized as a gauged sigma model. 
Using \eqref{Z-V}, this can be easily proved as follows. We have
 $$\mathcal{P}_{\g}(G,g_{0})=\mathcal{P}_{-\g}(G,g_{0})\cdot k,$$ 
 where the path $k(t)=\exp(\pi tT_3 /T)$ connects $-I$ and $I$ in $G$. Changing variables $g(t)\to g(t)k(t)$ in the path integral \eqref{A-path-integral} shifts $J\mapsto J+\pi T_3/T$, which can be compensated by the shift $h\rightarrow h+\pi$, which is allowed because the integrand is $2\pi$-periodic function of $h$.
 \end{remark}
 \begin{remark}\label{3}
 Assuming that the deformation $V_{I}$ is invariant under the shift $g(t) g$ with constant $g\in G$, we have from \eqref{A-path-integral},
$$ \mathcal A_{I} (g_0;\lambda )=\mathcal{A}_{I}(e,\lambda).$$
\end{remark}

Let $G_{\gamma}$ be the centralizer of $\gamma$ in $G$. `Unfolding' the fundamental domain $F=\Gamma\bk G$, we get
\beq
\sum_{\sigma \in\Gamma_\g \bk \Gamma } \int_F \mathcal A_{\sigma^{-1} \gamma \sigma } (g_0;\lambda )dg_0
&=\int_{\Gamma_\gamma\bk G} \mathcal A_{ \gamma  }  (g_0;\lambda )dg_0
\\ &=\int_{G_\gamma\bk G} dg_0  \int_{\Gamma_\gamma\bk G_\gamma} \mathcal A_{ \gamma  }  (g_1 g_0;\lambda )dg_1\\
&=\text{vol}(\Gamma_\gamma\bk G_\gamma)\int_{G_\gamma\bk G}   \mathcal A_{ \gamma  }  (g_0;\lambda )dg_0.
\eeq
As a result, we obtain the localized version of the pre-trace formula
\beq\label{Z-pretrace}
Z(iT)&=\sum_{[\gamma]}  \text{vol}(\Gamma_\gamma\bk G_\gamma) \int_{G_\gamma\bk G}dg_0{e^{-{i \ex{\rho,\rho}T \ov 2}}} \int _{-\pi}^{\pi} {{dh\ov 2\pi}\, } {i \ov 2\sin{ h }}   \mathcal A_{ \gamma  }  (g_0;\lambda )
\\&=\sum_{[\gamma]}  \text{vol}(\Gamma_\gamma\bk G_\gamma){e^{-{i \ex{\rho,\rho}T \ov 2}}} \int _{-\pi}^{\pi} {{dh\ov 2\pi}\, } {i \ov 2\sin{ h }} \int_{G_\gamma\bk G}   \mathcal A_{ \gamma  }  (g_0;\lambda )dg_0,
\eeq
where summation goes over conjugacy classes $[\ga]$ of the discrete subgroup $\Gamma$ of $G$, and in the 2nd line we interchanged the order of the integrations. According to Remarks \ref{2} and \ref{3}, contributions from the conjugacy classes $[\g]$ and $[-\g]$ to \eqref{Z-pretrace} are the same and for
$\g=I$ the integral over $G_{\gamma}\bk G$ is understood as   $\mathcal A_{I}  (e;\lambda )$. Also note that, as it  it follows from Remark \ref{measure},  the volume form $dg_0$ is given by formula \eqref{Haar} in Appendix \ref{A}. 

We can regard \eqref{Z-pretrace} as a path integral version of the pre-trace formula with an invariant deformation from the localization parametrized by $\lam$, where $\lam=0$ corresponds to the original pre-trace formula. The main role of the localization is that it is separately applied to each orbital integral over $G_{\ga}\bk G$, labeled by the conjugacy class $[\gamma]$, and upon taking the limit $\lambda\rightarrow \infty$  leads to  the desired Selberg trace formula.

\enlargethispage{3em}
\subsection{The Selberg trace formula}\label{sbc: STF} 
To make a precise connection with the Selberg trace formula,   we group the conjugacy classes in $\Gamma$ according to the conjugacy classes in $\bar\Gamma=\Gamma/\{\pm I\}\subset\mathrm{PSL}(2,\RR)$. Since $\bar\Gamma$ is cocompact, the  elements or $\bar\Gamma$ except the identity are either hyperbolic or elliptic. We have the following contributions to the pre-trace formula \eqref{Z-pretrace}.

\noindent
$\bullet$ {\it Identity element}
\medskip

For the conjugacy class $[I]=I$ in $\Ga$ we use the following invariant deformation
\beq
~& V_0=-{1\ov 2}\int_{0}^{T} \ex{\pa_t^{A_0} J,\pa_t^{A_0} \psi}dt,
\\& \del V_0={1\ov 2}\int_{0}^{T}(\ex{\pa_t^{A_0} J,\pa_t^{A_0} J}+i\ex{\pa_t^{A_0}\psi,\pa_t^J \pa_t^{A_0}\psi})dt.
\eeq
Here $A_0$ is some arbitrary regular element in $\mathfrak k$, introduced in order to have isolated critical points, say $A_{0}=2\pi T_{3}/T$. Then from the formula for $\del V_0$, we see that the path integral $\mathcal A_{ I}$ localizes onto a set of critical points
\beq
g_n(t)=g(0)e^{{2 \pi i \nu_n t \ov T} },\quad\text{where}\quad \nu_n=i n T_3,\quad n\in \mathbb Z\quad \text{and}\quad g(0)\in F.
\eeq
Indeed, equation $\pa^{A_{0}}_{t}J=0$ gives $J(t)=e^{-A_{0}t}J_{0}e^{A_{0}t}$, and since $J\in L\frak{g}$ we obtain $J_{0}=cA_{0}$; then $g(t)=g(0)e^{cA_{0}t}\in LG$ implies (for our choice of $A_{0}$) that $c$ is an integer. 

Around each critical point $g=g_n$ the leading quadratic term in $\del V_0$ has additional fermionic zero modes associated with the kernel of the operator $D=\pa_t^{J_n}$, where $J_{n}=g^{-1}_{n}\dot{g}_{n}=-2\pi n T_{3}/T$, which are given by
\beq
\text{Re}(E_{ \al} e^{{-2\pi i {\ex{\al,\nu_n}}t \ov T}}),\quad\text{Im}(E_{ \al} e^{{-2\pi i {\ex{\al,\nu_n}}t \ov T}})\quad \al\in R^{\mathfrak k}.
\eeq

To compute the contribution of these critical points, we use the method developed in \cite{Choi:2021yuz} (and refer to \cite[Sect. 5]{Choi:2021yuz} for more details). Namely,  denoting the fermion zero modes by $\chi$ and using the decomposition $\psi=\chi+\eta$, we see that 
\beq
\lim_{\lambda\to\infty}\mathcal{A}(e,\lambda)=\sum_{n\in\ZZ}e^{iS_{n}}\bm\int e^{i S_{n}^{\text{loc}}}\cD Y\cD\eta d\chi,
\eeq
where $S_{n}=S_f[\bm{h}](g_{n}(t),0)$ is the classical contribution to the gauge fixed action \eqref{eq:Sf}, $Y=g^{-1}\del g$ and
\beq
~&S^{\text{loc}}_{n}={1\ov 2}\int_{0}^{T}\left(\ex{ D\pa_t^{A_0}  Y, D\pa_t^{A_0}  Y}+i\ex{\chi,\pa_t^{A}\chi}+i\ex{\pa_t^{A_0}\eta, D \pa_t^{A_0} \eta}
\right.\\&\left.+i\ex{\pa_t^{A_0}\eta, [DY,\pa_t^{A_0}\chi]}+{i \ov 2}\ex{\pa_t^{A_0}\chi, [[DY,Y],\pa_t^{A_0}\chi]}\right)dt.
\eeq
Since $[Y,\pa_t^{A_0}\chi]$ is orthogonal to $\chi$, the change of variables 
\beq
\pa_t^{A_0}\eta \rightarrow \pa_t^{A_0} \eta-[Y,\pa_t^{A_0}\chi]
\eeq
is legitimate, and we obtain
\beq
S_{n}^{\text{loc}}={1\ov 2}\int_{0}^{T}\left(\ex{ D\pa_t^{A_0}  Y, D\pa_t^{A_0}  Y} + i\ex{\chi,\pa_t^{A}\chi}+i\ex{\pa_t^{A_0}\eta, D \pa_t^{A_0} \eta}\right)dt.
\eeq

Evaluating Gaussian path integrals, we finally get \beq
  \mathcal A_{ I }  (g_0;\lambda )&={1\ov (2\pi i) (-2\pi  i )^{1/2}}\frac{\mathrm{Pf}(-D(\pa_{t}^{A_{0}})^{2})}{\sqrt{\det D^{2}(\pa_{t}^{A_{0}})^{2}}}\bm\int e^{ -{1\ov 2} \int_{0}^{T}\ex{\chi,\pa_t^{A} \chi} dt}d\chi
  \\&={1\ov (2\pi i) (-2\pi  i )^{1/2}} \cdot {1\ov  i T\cdot T^{1/2} } \cdot 2(h+2\pi n)
  \\&={2(h+2\pi n) \ov (2\pi iT)^{3/2} } ,
  \eeq
  where we used Remark \ref{remark:dim} and section 4 of \cite{Choi:2021yuz} for the computations of each factors. 
  
Now we use the important relation, valid whenever $-I\in \Gamma$, which holds for any Riemannian metric on $G$ and corresponding induced Riemannian volumes on $\Gamma\bk G$ and $K$,
\beq \label{eq:volumerelation}
\text{vol}(\Gamma \bk G)/\text{vol}(K)=\text{vol}(X)/2.
\eeq
Indeed, $X=\Gamma\bk G/K$ and $\Gamma\cap K=\{\pm I\}$, so $\bar{K}=K/\{\pm I\}$ acts freely on $\Gamma\bk G$.
We emphasize that in our case the Riemannian volume form is induced from Cartan-Killing metric $ds^2=4\Tr(g^{-1}dg\, g^{-1}dg)$ (see Appendix \ref{A}), so we obtain
\beq \label{eq:trivloc}
Z_{I}(iT) ={\text{vol}(X) e^{-{i T\ov 16}}\ov 2(\pi iT)^{3/2}}\int_{-\pi}^\pi \sum_{n\in\mathbb Z} {i(h+2\pi n)\ov \sin h}e^{-{4i (h+2\pi n)^2\ov T}}dh.
\eeq 

\begin{remark} We emphasize that the infinite series in \eqref{eq:trivloc} is well defined, since the apparent poles at $h=m\pi$ cancel each other once we group together terms with $n$ and $-n-m$. This makes sense, since according to the formula \eqref{prop-on-G}, the series \eqref{eq:trivloc} is a propagator on $\mathrm{SL}(2,\RR)$.
\end{remark}
However, in order to Wick rotate expression \eqref{eq:trivloc} to the Euclidean time we need to incorporate the integration over $h$. Naively changing the order of integration and summation,  we arrive at a remarkably simple formula
\beq 
Z_{I}(iT)&\stackrel{\text{naive}}{=}{\text{vol}(X)e^{-{i T\ov 16}}\ov  2(\pi iT)^{3/2}}\int_{\mathbb R} {ih\ov \sin h}e^{-{4i h^2\ov T}}dh.
\eeq

However, this representation is ambiguous because of the poles at $h=n\pi$ with non-zero integer $n$, and we need to specify a precise choice of the contour of integration that bypasses these poles.

We claim that there is a unique choice of such contour which gives a physically meaningful thermal partition function after the Wick rotation to the Euclidean time. It resembles the Feynman contour, and we write
\beq \label{eq:trivlorentzfinal}
Z_{I}(iT)&\stackrel{\text{physical}}{=}{\text{vol}(X)e^{-{i T\ov 16}}\ov2 (\pi iT)^{3/2}}\int_{\mathbb R(1-i\varepsilon)} {ih\ov \sin h}e^{-{4i h^2\ov T}}dh.
\eeq

It is quite remarkable that after the Wick rotation with $T=-i\beta$ for this choice of the contour of integration, twice the contribution of $\gamma=I$ is equal to the contribution of the identity element to the Selberg trace formula,
\beq \label{eq:ztriv}
2Z_{I}(\be) &={\text{vol}(X)e^{-{\be\ov 16}}\ov (\pi \be)^{3/2}} \int_{-\infty}^{\infty} {p\ov \sinh p} e^{{-}{ 4p^2\ov \be}}dp .
\eeq

We argue that the any other choice of the contour gives the physically meaningless partition function. Indeed, suppose we choose another contour that bypasses the singularities at $h=\pi n$ differently. Then at each pole where this contour differs from the Feynman like contour,  \eqref{eq:trivlorentzfinal} gets an additional contribution of the form $e^{-i C/T}$ with $C>0$. However, this term analytically continues to $e^{C/\be}$ for $\text{Im}\, T=-\beta<0$, and this exponential high temperature growth contradicts the Weyl's law \cite{weyl1911asymptotische}, which states that  $Z(\be)\sim \be^{-d/2}$ as $\be\to 0$.

\medskip
\noindent

$\bullet$ {\it Hyperbolic elements}
\medskip

Element $\g\in\Ga$ is hyperbolic, if $|\Tr\ga|>2$, and since $Z_{[\ga]}=Z_{[-\ga]}$, we can always choose $\ga$ such that $\Tr\ga>2$. Every such $\ga$ is conjugated in $G$ to the element $e^{rT_{1}}$, where
$r=\log\lambda(\ga)$ and $\lambda(\ga)$ is the multiplier of $\ga$. We have $\Ga_{\ga}=\Ga_{\ga_{0}}$ and $G_{\ga}=G_{\ga_{0}}$, where $\ga_{0}$ is a primitive hyperbolic element, $\ga=\ga_{0}^{k}$
for $k\geq 1$. 
Thus we can choose $\gamma=e^{rT_1}$ and $\gamma_0=e^{r_0T_1}$ so $G_\gamma=\{ e^{aT_1}| \, a\in \mathbb R \}\cup \{ -e^{aT_1}| \, a\in \mathbb R \} $. Choosing the measure on $G_\gamma$ to be $da$, we have
$$\text{vol}(\Gamma_{\gamma}\bk G_{\gamma})=\text{vol}(\Gamma_{\gamma_{0}}\bk G_{\gamma_{0}})=r_{0},\quad\text{where}\quad r_{0}=\log\lambda(\ga_{0}).$$
and
$$ Z_{[\ga]}=r_{0}{e^{-{i \ex{\rho,\rho}T \ov 2}}} \int _{-\pi}^{\pi} {{dh\ov 2\pi}\, } {i \ov 2\sin{ h }} \int_{G_{\gamma}\bk G}   \mathcal A_{ \gamma  }  (g_0;\lambda )dg_0,$$
where the measure on $G_{\gamma }\bk G$ should reflect our choice of the measure on $G_{\gamma }$ above and will be determined later. Note that the final result is independent of this choice since $G=G_\gamma \times ( G_\gamma\bk G )$.

Now we compute $Z_{[\ga]}$ using the following deformation
\beq
~& V=-{1\ov 2}\int_{0}^{T}\ex{\pa_t J,\pa_t \psi} dt,
\\& \del V={1\ov 2}\int_{0}^{T} (\ex{\pa_t J,\pa_t J}+i\ex{\pa_t\psi,\pa_t^J \pa_t\psi})dt.
\eeq

The stationary points are determined from the equations $\dot J=0$ and $\pa_{t}D\dot\psi=0$, where $D=\pa_{t}^{J}$, so 
\beq
g_\gamma(t)=e^{\bm r t/T}g_0 \quad\text{and}\quad J_\gamma(t)=\frac{1}{T}\text{Ad}_{g_0^{-1}} \bm r,
\eeq
where $\bm{r}=rT_{1}$ and $\psi=\psi_{0}=\sum_{a=1,2}\psi^{a}_{0}T_{a}\in\Pi\frak{p}$.
As in the previous case $\gamma=I$, we easily compute the finite-dimensional fermionic integral 
\beq\label{chi-finite-dim}
\int e^{-\frac{h}{2}\la\psi_{0},[T_{3},\psi_{0}]\ra}d\psi_{0}=2h,
\eeq
and the remaining Gaussian path integral from the localization
\begin{equation}\label{Gauss-hyper}
\begin{split}
&\bm\int e^{\frac{i}{2}\int_{0}^{T}(\ex{ D\pa_t  Y, D\pa_t Y}+i\ex{\pa_t\eta, D \pa_t\eta})}\cD Y\cD\eta= \frac{1}{(2\pi i)(-2\pi i)^{1/2}}\frac{\text{Pf}(-\pa^{2}_{t}D)}{\sqrt{\det \pa^{2}_{t} D^{2}}}  
\\
&\hspace{10mm}=\frac{r}{(2\pi iT)^{3/2}\sinh r}.
\end{split}
\end{equation}
Thus we obtain
\beq \label{eq:zhyp}
\begin{split}
&Z_{[\gamma]}(iT)=\frac{r_{0}re^{-\frac{i T}{16}}}{(2\pi iT)^{3/2}\sinh r} \int _{-\pi}^\pi{dh\ov 2\pi}{ ih \ov \sin h }  \int \limits _{G_{\gamma}\bk G}e^{iS[\bm{h}](g_\gamma,0)} dg_{0}\\
 &
 =\frac{r_{0}re^{\frac{4ir^{2}}{T}-\frac{i T}{16}}}{(2\pi iT)^{3/2}\sinh r}\int_{-\pi}^\pi {dh\ov 2\pi}  {ih e^{\frac{4ih^{2}}{T}}\ov \sin h} \int_{G_{\gamma}\bk G}  e^{-{i \ex{\Ad{g_0}\bm h ,\bm r}\ov T}}dg_{0}
  \\
  &=\frac{2r_{0}re^{\frac{4ir^{2}}{T}-\frac{i T}{16}}}{(\pi iT)^{3/2}\sinh r}\int_{-\pi}^\pi dh  {i h e^{\frac{4ih^{2}}{T}}\ov \sin h}\int_{-\infty}^{\infty} e^{{8ihr n\ov T}}dn
 \\
 & = \frac{1}{2}\left({1\ov \pi iT}\right)^{1/2} { r_0 \ov \sinh {r}}e^{{4i r^2\ov  T}-{i T\ov 16}  }.
 \end{split}
\eeq

Here the orbital integral over $G_\gamma\bk G$ in the third equality is computed as following. In addition to the Iwasawa decomposition $G=ANK$ (see Appendix \ref{A}), we use the following parametrization of a generic element $g\in SL(2,\mathbb R)$,
\beq 
g=\begin{pmatrix} e^a & 0 \\ 0 & e^{-a} \end{pmatrix} \begin{pmatrix} 1 & n \\ 0 & 1 \end{pmatrix} \begin{pmatrix}\;\;\cos\theta & \sin\theta\\
\!-\sin\theta &\cos\theta\end{pmatrix}, \quad  a,n\in \mathbb R, ~ 0<\theta\leq 2\pi,
\eeq
with the measure $dg=8\sqrt{2}dadnd\theta $ is induced from the Cartan-Killing metric $ds^2=4\Tr(g^{-1}dg \,g^{-1}dg)$. Since  $G_{\ga}=\{ e^{aT_1}| \, a\in \mathbb R \}\cup \{ -e^{aT_1}| \, a\in \mathbb R \} $, we have a natural parametrization of $G_\gamma\bk G$ as
\beq
G_\gamma\bk G&=\mathbb Z_2 \bk NK
\\&=\left\{  \begin{pmatrix} 1 & n \\ 0 & 1 \end{pmatrix} \begin{pmatrix}\;\;\cos\theta & \sin\theta\\
\!-\sin\theta &\cos\theta\end{pmatrix}, \quad  a,n\in \mathbb R, ~ 0\leq \theta< \pi,  \right\}.
\eeq
The measure $da$ on $G_\gamma$ fixes the measure on $G_\gamma \bk G$ to be $dg_0=8\sqrt{2}dndk$, where $n\in \mathbb R$ and $0\leq \theta <\pi$. Therefore, the orbital integral becomes
\beq
\int_{G_{\gamma}\bk G}  e^{-{i \ex{\Ad{g_0}\bm h ,\bm r}\ov T}}dg_{0}=8\pi \sqrt{2}\int_{-\infty}^{\infty} e^{{8ihr n\ov T}}dn=\frac{2\sqrt{2}\,T}{r}\delta(h).
\eeq

Finally, the last expression \eqref{eq:zhyp} of can be analytically continued to $T=-i\beta$ and we obtain
\beq
Z_{[\gamma]}(\be)= {1 \ov 2\sqrt{\pi \be}} { r_0 \ov \sinh r}e^{-{4 r^2\ov \be}-{\be \ov 16}}.
\eeq

\medskip
\noindent
$\bullet$ {\it Elliptic elements}
\medskip

An elliptic element  $\gamma$ can written it as $\gamma=e^{\bm \theta}$ with $\bm \theta=\theta \tilde T_3$, where $\tilde T_3$ is conjugate to $T_3$ in $G$, and $\theta\in (0,\pi) \cup (\pi,2\pi)$. Again, since the intervals $(0,\pi)$ and $(\pi,2\pi)$ are related by $\gamma=-I$, it is sufficient to compute the case $\theta \in (0,\pi)$ and to use same deformation as in the hyperbolic case,
\beq
~& V=-\int dt {1\ov 2}\ex{\pa_t J,\pa_t \psi}
\\& \del V=\int dt {1\ov 2}\ex{\pa_t J,\pa_t J}+{i\ov 2}\ex{\pa_t\psi,\pa_t^J \pa_t\psi}.
\eeq
This gives the set of isolated critical points 
\beq
g_{n}(t)=e^{\bm \theta_n t/T}g_0 \quad \text{and}\quad J_n(t)=\text{Ad}_{g_0^{-1}}\bm \theta_n,
\eeq
where 
$$\quad \bm\theta_n=\theta_{n}\tilde T_3\quad\text{and}\quad \theta_{n}=\theta+2\pi n,\quad n\in\ZZ.$$
Their contribution to the partition function is given by
\begin{equation}
\begin{gathered} 
Z_{[\gamma]}(iT)\\
 ={\text{vol}(\Gamma_\gamma \bk G_{\gamma})e^{-{i T\ov 16}} \ov (2\pi i T)^{3/2}} \int_{-\pi}^\pi {dh\ov 2\pi}\sum_{n\in \mathbb Z}  \int \limits_{\,G_{\gamma}\bk G}dg_0  {ih \theta_n \ov \sin h \sin\theta_n}e^{-{4i( \theta_n^2+h^2)+i \ex{\Ad{g_0}  \bm h , \bm \theta_n}\ov T}}.
\end{gathered}
\end{equation}

Here we've been careful with the order of integration and summation. The orbital integral over $G_\gamma\bk G$ can computed  explicitly using the Cartan decomposition $G= KAK$,
\beq \label{eq:KAK}
g=\begin{pmatrix}  \,\,\,\cos \al & \sin \al \\ \!-\sin\al & \cos\al \end{pmatrix}  \begin{pmatrix}  e^t &0\\ 0 & e^{-t} \end{pmatrix}
\begin{pmatrix}  \,\,\,\cos w & \sin w \\\! -\sin w & \cos  w \end{pmatrix},
\eeq
where $t\geq 0$, $0\leq \al<2\pi$ and $0 \leq w <\pi$.
Here the $dg$ measure is induced from the Cartan-Killing metric $ds^2=4\Tr(g^{-1}dg \,g^{-1}dg)$ and we have 
$$dg=16\sqrt{2} \sinh(2t)dtd\al dw.$$ 

Since $G_\gamma=K$, the natural parametrization of $g_0\in G_\gamma\bk G$ is
\beq
g_0= \begin{pmatrix}  e^t &0\\ 0 & e^{-t} \end{pmatrix}
\begin{pmatrix}  \,\,\,\cos w & \sin w \\ \!-\sin w& \cos w \end{pmatrix},\quad t\in\mathbb R ,\quad 0\leq w<\pi.
\eeq
As in the hyperbolic case,  we choose the measure on $G_\gamma$ to be $d\al$, so
\beq
\text{vol}(\Gamma_\gamma\bk  G_\gamma)=\theta_p, \quad dg_0= 16\sqrt{2}\sinh (2t) dt d w,
\eeq
where $\theta_p$ is such that $e^{\theta_p T_3}$ is the primitive element in $\Gamma_\gamma$. As before, the measure on $ G_\gamma\bk G$ is determined from $G=G_\gamma \times ( G_\gamma\bk G )$.

The relevant orbital integral becomes
\beq
\int_{G_\gamma \bk G}dg_0e^{-{i \ex{\Ad{g_0}  \bm h , \bm \theta_n}\ov T}}&=16\pi\sqrt{2}\int_0^\infty  e^{{8i h\theta_n \cosh(2t)\ov T}} \sinh (2t) dt
\\&=8\pi \sqrt{2} \int_1^\infty  e^{8ih \theta_n y\ov T} dy 
\\&={i\pi \sqrt{2}T \ov h\theta_n}e^{8ih\theta_n\ov T} +\sqrt{2}\pi^2 T \del(h\theta_n),
\eeq
 which gives 
  \begin{equation*}
  \begin{gathered}
Z_{[\gamma]}(iT)\\
={\theta_pe^{-{i T\ov 16}}\ov 2(2\pi i T)^{3/2}}  \int_{-\pi}^\pi  dh  \sum_{n\in \mathbb Z}  {ih\ov \sin h}{\theta_n\ov \sin {\theta_n}}e^{-{4i( \theta_n^2+h^2)\ov T}}  \left({i\sqrt{2}T\ov h \theta_n }e^{{8ih\theta_n \ov T}} +{\sqrt{2} \pi T\ov \theta_n} \del(h)\right)
\\={\theta_pe^{-{i T\ov 16}} \ov 4(\pi i T)^{1/2}\sin\theta}  \int_{-\pi}^\pi  dh  \sum_{n\in \mathbb Z}  {ih\ov \pi\sin h}e^{-{4i( \theta_n-h)^2\ov T}}  {1\ov h+i0_+}.
\end{gathered}
\end{equation*}
To  simplify this expression, we formally interchange the integration over $h$ and the summation over the set of critical points. As before, we choose the Feynman-like contour which is uniquely determined by the  compatibility with the Weyl's law. Such contour for  $\theta\in(0,\pi)$ is $ \mathcal C=\mathbb R(1+i\ep)$, and we obtain
 \beq 
 Z_{[\gamma]   }(iT)={\theta_p e^{-{i T\ov 16}}\ov 4(\pi i T)^{1/2}\sin\theta}  \int_{\mathcal C}   {i\ov \pi\sin h}e^{-{4i( \theta-h)^2\ov T}} dh.
\eeq
The analytically continued expression can be compactly written as
 \beq 
Z_{[\gamma]}(\be)&={\theta_p e^{-{\be\ov 16}}\ov 4(\pi \be)^{1/2}\sin \theta}  \left(  \text{v.p.} \int_{\mathbb R}   {i\ov \pi\sinh p}e^{-{4( p+i\theta)^2\ov \be}} dp +e^{{4 \theta^2\ov \be}}\right).
\eeq

We can simplify further by representing the exponent  $e^{-{4( p+i\theta)^2\ov \be}}$ as another Gaussian integral and changing the order of integrations (which is now legitimate)
\begin{equation*}
\begin{gathered} 
Z_{[\gamma]}(\be)\\={\theta_p e^{-{\be\ov 16}}\ov 4(\pi \be)^{1/2}\sin\theta} \left( \text{v.p.}{i\ov \pi}\sqrt{{\be\ov 16\pi}} \int_{\mathbb R} dp \int_\mathbb R \, {e^{-{\be u^2\ov 16}} \cos ((p+i\theta) u)\ov \sinh{p}}  du
+e^{4\theta^2/\be} \right)
  \\={\theta_p e^{-{ \be\ov 16}} \ov 16\pi \sin\theta}\left({ 1\ov \pi} \int_{\mathbb R} du\int_{\mathbb R} \, e^{-{\be u^2\ov 16}}{\sin(pu)\sinh(\theta u)\ov \sinh{p}}dp  
+\int_\mathbb R du \, e^{-{\be u^2\ov 16}} \cosh (\theta u)
 \right)
      \\={\theta_p e^{-{\be\ov 16}} \ov 16\pi \sin\theta}\int_{\mathbb R} du\, e^{-{\be u^2\ov 16}} {\cosh({(\pi/2-\theta)u})\ov \cosh(\pi u/2)}.
\end{gathered}
\end{equation*}

\medskip
\noindent
$\bullet$ {\it Final Result}
\medskip

Collecting results in this section we obtain the Selberg trace formula for the compact Riemann surface $X=\Ga\bk\HH$,
\begin{equation}
\begin{gathered}
\Tr_{L^2_0(X)}[e^{-\be \Delta/2}]={\text{vol}(X)e^{-{ \be\ov 16}}\ov (\pi \be)^{3/2}} \int_{\mathbb R} dp\, {p\ov \sinh p} e^{{-}{ 4p^2\ov \be}}
\\+\sum \limits_{\substack { [\gamma]~ \text{hyperbolic}\\ \Tr[\gamma]>2}}   {1 \ov \sqrt{\pi \be}} { r_0 \ov \sinh {r}}e^{-{4 r^2\ov \be}-{\be \ov 16}} 
\\+\sum_{ \substack{ [\gamma] ~\text{elliptic} \\ 0<\theta<\pi}} {\theta_p \ov 8\pi \sin\theta}\int_{\mathbb R} du\, e^{-{\be (u^2+1)\ov 16}} {\cosh({(\pi/2-\theta)u})\ov \cosh(\pi u/2)}.
\end{gathered}
\end{equation}
After the rescaling $\Delta\rightarrow \Delta/2$ and  $\be \rightarrow 4\be$, it exactly matches  the standard form of the Selberg trace formula \cite{selberg1956harmonic, MR0439755} 
\begin{equation}
\begin{gathered}
\Tr_{L^2_0(X)}[e^{-\be \Delta}]={\mu(X) e^{-{\be \ov 4}}\ov (4\pi \be)^{3/2}} \int_0^\infty dp\, {p\ov \sinh {p\ov 2}} e^{{-}{ p^2\ov 4\be}}
\\+\sum_{\substack { [\gamma]~ \text{hyperbolic}\\ \Tr[\gamma]>2}}  {1 \ov \sqrt{4\pi \be}} { r_0 \ov \sinh {r}}e^{-{ r^2\ov \be}-{\be \ov 4}} 
\\+\sum_{ \substack{ [\gamma] ~\text{elliptic} \\ 0<\theta<\pi}} {\theta_p\ov 4\pi \sin\theta}\int_{\mathbb R}{\cosh((\pi-2\theta)r)\ov \cosh(\pi r)}e^{-\be (r^2+{1\ov 4})}dr,
\end{gathered}
\end{equation}
where $\mu(X)=\text{vol}(X)/2$ is the hyperbolic area of $X$.

\subsection{Generalization to arbitrary representations} \label{sec:genrep}

The trace formula on the Hilbert space $L_{0}^2(X)$ can be generalized to a trace formula on the  Hilbert space $L_{0}^2(X;\pi)$ associated with a unitary finite-dimensional representation $\pi$\footnote{There should be no confusion between the representation $\pi$ and the area of the circle of radius $1$.} of the Fuchsian group $\Gamma$. Elements $\Psi\in L_{0}^2(X;\\\pi)$ can be thought as sections of a vector-bundle on $X$ associated with the representation $\pi$ or, equivalently, as vector-valued functions  $\Psi$ on $G$, satisfying the following constraints
\beq
\Psi(\gamma x)=\pi(\gamma)\Psi(x), \quad\Psi(xk)=\Psi(x), \quad  x\in G, \gamma\in \Gamma, k \in K.
\eeq

To perform a path integral quantization,  we lift the representation $\pi$ of $\Gamma$ to the representation of $\pi_1(\Gamma\bk G,\bar e)$ and observe that for each connected component of the domain of integration \eqref{main-decomposition} we get an additional weight factor $\chi_{\pi}(\gamma)$, where $\chi_{\pi}$ is the character of $\pi$ (cf. \cite{Schulman:1968yv, Laidlaw:1970ei, Dowker:1972np}).
Hence the partition function decomposition \eqref{eq:naturaldecomp} is modified as
\beq
Z(iT;\pi)=\sum_{[\gamma] } \chi_\pi (\gamma) Z_{[\gamma]} (iT),
\eeq
where 
$Z_{[\gamma]}(iT)$ is given by the same path integral that we previously localized in the case of trivial representation. Consequently, we obtain
the trace formula on $L^2(X;\pi)$,
\begin{equation}
\begin{gathered} 
\Tr_{L^2(X;\pi)}[e^{-\be \Delta/2}]={d_{\pi }\text{vol}(X)e^{-{ \be\ov 16}}\ov (\pi \be)^{3/2}} \int_{\mathbb R} dp\, {p\ov \sinh p} e^{{-}{ 4p^2\ov \be}}
\\+\sum_{\substack { [\gamma]~ \text{hyperbolic}\\ \Tr[\gamma]>2}}{\chi_{\pi}(\gamma) \ov \sqrt{\pi \be}} { r_0 \ov \sinh {r}}e^{-{4 r^2\ov \be}-{\be \ov 16}} 
\\+\sum_{ \substack{ [\gamma] ~\text{elliptic} \\ 0<\theta<\pi}}  {\chi_{\pi}(\gamma)\theta_p \ov 8\pi \sin\theta}\int_{\mathbb R} du\, e^{-{\be (u^2+1)\ov 16}} {\cosh({(\pi/2-\theta)u})\ov \cosh(\pi u/2)},
\end{gathered}
\end{equation}
which simply reproduces the standard result \cite{selberg1956harmonic, MR0439755}.
\subsection{Generalization to arbitrary weights}

It is quite amusing to observe that the path integral derivation of the Selberg trace formula provides a unified simple approach for the case of automorphic forms of arbitrary weight $m\in {1\ov 2}\mathbb Z$.

Let us first consider the case of integer weight $m\in \mathbb Z$.  As in the case of automorphic functions ( the forms of zero weight), one can show the following equality
\beq 
\Tr_{L^2_m(X)}[e^{-iT \Delta/2}]={e^{-{i \ex{\rho,\rho}T \ov 2}}\ov\text{vol}(\mathcal G)} \bm{\int}  {\mathcal W_{2m}(A) \ov \mathcal W_{-1}(A)-\mathcal W_1(A)} \psi_{\mathfrak k}e^{i S_{A}[g,\psi]}\mathscr Dg\mathscr  D\psi \mathscr  DA.
\eeq
The only difference with the case $m=0$ is the appearance of the additional Wilson line in the path integral, which depends only on the gauge field. Therefore, the localization procedure is almost identical to the weight zero case with the same prescription for the physical contour! 
Thus after the Wick rotation we obtain
\begin{equation} \label{eq:arbitrary weight}
\begin{gathered}
\Tr_{L^2_m(X)}[e^{-\be \Delta/2}]={\text{vol}(X)e^{-{ \be\ov 16}}\ov (\pi \be)^{3/2}} \int_{\mathbb R} dp\, {p \cosh(2mp)\ov \sinh p} e^{{-}{ 4p^2\ov \be}}
\\+\sum_{\substack { [\gamma]~ \text{hyperbolic}\\ \Tr[\gamma]>2}} {1 \ov \sqrt{\pi \be}} { r_0 \ov \sinh {r}}e^{-{4 r^2\ov \be}-{\be \ov 16}} 
\\+\sum_{ \substack{ [\gamma] ~\text{elliptic} \\ 0<\theta<\pi}}  {\theta_p e^{-{\be\ov 16}}\ov 2(\pi \be)^{1/2}\sin\theta}  \left(  \int_{\mathbb R}   {i e^{2mp}\ov \pi\sinh p}e^{-{4( p+i\theta)^2\ov \be}} dp +e^{{4 \theta^2\ov \be}}\right),
\end{gathered}
\end{equation}
As in Section  \ref{sec:genrep}, this formula admits a straightforward generalization to the Hilbert space $L^2_m(X;\pi)$ associated with a general unitary finite-dimensional representation  $\pi$ of $\Gamma$ satisfying $\pi(-I)=\pi(I)$.

However, in case of $m\in \mathbb Z+{1\ov 2}$ we have seen in Section that  
the Hilbert space $L^2_m(X)$ is empty because $-1\in \Gamma$. Therefore, it is necessary to consider a non-trivial one-dimensional representation $\pi$ which obeys the consistency condition $\pi(-I)= -1$. 
It is striking to see how simply the path integral approach reproduces the correct result! Thus performing the same manipulation as in $m=0$ case, in basic case $m=\frac{1}{2}$ we obtain the following formula
\begin{equation}
\begin{gathered}
\Tr_{L^2_{\frac{1}{2}}(X;\pi )}[e^{-\be \Delta/2}]={\text{vol}(X)e^{-{ \be\ov 16}}\ov 8\pi^2} \int_{\mathbb R} dp\, {p \coth p} e^{{-}{\be u^2  \ov 4}}
\\+\sum_{\substack { [\gamma]~ \text{hyperbolic}\\ \Tr[\gamma]>2}} {\chi_\pi(\gamma) \ov \sqrt{\pi \be}} { r_0 \ov \sinh {r}}e^{-{4 r^2\ov \be}-{\be \ov 16}} 
\\+\sum_{ \substack{ [\gamma] ~\text{elliptic} \\ 0<\theta<\pi}}  {\chi_\pi(\gamma)\theta_p \ov 8\pi \sin\theta} \left( \int_{\mathbb R}  e^{-{\be (u^2+1)\ov 16}} {\sinh({(\pi/2-\theta)u})\ov \sinh(\pi u/2)}du +2i\right).
\end{gathered}
\end{equation}
Needless to say that it  exactly matches the known result \cite{selberg1956harmonic, MR0439755}.

\begin{remark} The path integral approach we developed here  is also applicable to compact homogeneous spaces $G/K$, where
$G$ is a compact Lie group and $K$ is its subgroup.
In this case we can directly  
use the Euclidean formalism for formulas like \eqref{eq:mainint}, which is a major simplification. This approach provides a compact and simple derivation of known results related to the Eskin trace formula, including the basic case $\mathrm{SU}(2)/\mathrm{U}(1)\simeq S^2$
(see \cite{Camporesi:1990wm} and references therein).
\end{remark}

\section{Selberg trace formula on general compact locally symmetric space} \label{sec:general}

Here we generalize further our path integral localization approach to include 
the Selberg trace formula for a general compact locally symmetric space $X=\Gamma\bk G/K$, where $G$ is a non-compact real semi-simple Lie group of rank $r_{G}$, $K$ is its maximal compact subgroup and $\Ga$ is a discrete subgroup of $G$ such that $\Ga\bk G$ is compact. For simplicity, we consider only  the trace formula for Laplace operator acting on functions  on $X$, which is enough to convey the main novelty.

\subsection{General set-up for $\Ga\bk G/K$}\label{s:5.1}
We start with the same gauged sigma model \eqref{Lagrangian} on $G/K$ with the same supersymmetry \eqref{eq:susyselberg}, where now $A$ is a connection in a principal $K$-bundle over $S^{1}_{T}$. There is a main technical difficulty if one tries to follow the logic developed for the Riemann surfaces due to the fact that the structure group $K$ is non-abelian and hence the Gauss law (with or without additional Wilson lines) would be more complicated. Namely, it is difficult to anticipate the analogous insertion of Wilson lines to the path integral that gives the Hilbert space $L^{2}_{0}(X)$.

It is remarkable that there is a short-cut, which bypasses the above obstacles and produces the desired path integral identity for the trace formula! 

Namely, let $P$ be the principal $G$-bundle over $S^{1}_{T}$ and let $\nabla_A=d+\text{ad}_A$ be a connection on the associated adjoint bundle $P\times_{\text{Ad}|_K}\, \mathfrak g$, where $\text{Ad}|_K:K\hookrightarrow{} G\rightarrow\text{Ad}(\mathfrak g)\subset \mathrm{GL}(\mathfrak g)$ is a natural adjoint action of $K$ on $\mathfrak g$. 
We claim that the generalization of \eqref{eq:maineq} is the following formula
\beq \label{Tr-general}
\begin{gathered}
\Tr[e^{-iT \Delta/2}]\\={e^{-i\ex{\rho,\rho}T\ov 2}\ov \text{vol}(\mathcal G)} \bm\int {\chi(A) e^{i \int_0^T ({1\ov 2}\ex{J_{A},J_{A}}+{i\ov 2}\ex{\psi,\pa_t^{A}\psi})dt}
\ov \text{Pf}{\,'} \left( i(\text{Hol}^{-1/2}_{S^1_T}(\nabla_A)-\text{Hol}^{1/2}_{S^1_T}(\nabla_A)) \right) }\mathscr D g\mathscr D\psi \mathscr DA .
\end{gathered}
 \eeq
Here $\text{Hol}_{S^1_T}(\nabla_A)\in \mathrm{GL}(\mathfrak g)$ is the the holonomy along the temporal circle and $\text{Pf}'$ is a Pfaffian on the non-zero eigenvalue subspace in $\mathfrak g$, which will become clear in a moment. Finally, $\chi(A)$ denotes the insertion of $r_{G}$ fermionic zero modes, associated with the kernel of the operator $\nabla_A$ acting on $\mathfrak g$-valued functions on $S^{1}_{T}$. 

Though the explicit form of $\chi(A)$ for generial time-dependent $A$ is complicated, it can be obtained by first fixing the gauge with $ A=\text{const}$ where the fermionic zero modes are simple, and then gauge transforming it back to get fermionic zero modes for generial  $A$. 

Namely, since two connections with the same holonomy $u\in K$ are gauge equivalent and the exponential map $\exp: \frak{k}\to K$ is surjective, for a connection $\nabla_{A}$ with the holonomy $u=e^{\bm h}$ where $\bm h \in \mathfrak k$, we can  
choose the gauge  $\dot A=0$, so $A=\bm h/T$. Since the conjugation by $K$ is a residual gauge symmetry, we can let $\bm h\in\mathfrak t$, where $\mathfrak t$ is the Cartan subalgebra of $\mathfrak k$. 
Since generic $A$ is regular in $\mathfrak k$, so that its centralizer in $\mathfrak k$ is $\mathfrak t$, i.e. $Z_{\mathfrak k}(A)=\mathfrak t$ and so $\text{ker}(\nabla_A)\simeq Z_{\mathfrak g}(\mathfrak t)$. Now it follows from \cite[Proposition 6.60]{knapp1996lie} that
\beq
Z_{\mathfrak g}(\mathfrak t)=\mathfrak h^c, 
\eeq
where $\mathfrak h^c\subset \mathfrak g$ is a maximally compact Cartan subalgebra in $\mathfrak g$ which is stable under the Cartan involution $\theta$ on $\mathfrak g$, which satisfies $\theta^2=1$ with $\mathfrak k$ and $\mathfrak p$ being $+1$ and $-1$ eigenspaces. This leads to a naturally decomposition $\mathfrak h^c=\mathfrak t\oplus \mathfrak a$, where $\mathfrak t\subset \mathfrak k$ and $\mathfrak a\subset \mathfrak p$.

This clearly shows that for general $A$ there are $r_G=\dim\mathfrak h^c $ linear independent fermionic zero modes. Representing them by the elements $\chi^{1},\dots ,\chi^{r_G}$ associated with the orthonormal basis of $\mathfrak h^c$, we have insertion of zero modes
\beq
\chi(A)=c_{r_G} \chi^{1}\cdots \chi^{r_G}.
\eeq

As in Section \ref{W-defects}, to prove the main formula  \eqref{Tr-general} it is sufficient to show that, with some appropriate normalization $c_G$,  the fermionic path integral cancels the denominator in \eqref{Tr-general}, which is the following identity
\beq \label{eq:ferintgen}
\bm \int  \chi(A)e^{i \int_0^T {i\ov 2}\ex{\psi,\pa_t^{A}\psi}dt} \mathscr D\psi =\text{Pf} {\, '}\left( i( \text{Hol}^{-1/2}_{S^1_T}(\nabla_A)-\text{Hol}^{1/2}_{S^1_T}(\nabla_A) )\right) .
\eeq

To prove \eqref{eq:ferintgen}, we again fix the gauge to $A=\bm h/T$ with $\bm h\in \mathfrak t$ and as in Section \ref{W-defects},  use the Hamiltonian formalism which generalizes \eqref{Wilson-fermi}.

Let $\mathfrak{h}^c _{\mathbb C}$ the complexification of $\mathfrak{h}^c$ which is a Cartan subalgebra of $\mathfrak g_{\mathbb C}$, which leads to a decomposition $\mathfrak g_{\mathbb C}=\mathfrak h^c_{\mathbb C} \oplus \bigoplus_{\al\in R_{\frak{g}_{\mathbb C},\frak{h}^c_{\mathbb C}}} \mathfrak g_\al$ where $R_{\frak{g}_{\mathbb C},\frak{h}^c_{\mathbb C}}$ is a root system of $\mathfrak g_{\mathbb C}$ w.r.t. $\mathfrak h^c_{\mathbb C}$. As in the Appendix A of \cite{Choi:2021yuz}, we introduce a Cartan-Weyl basis of $\mathfrak g_{\mathbb C}$ such that $\{H_i\}$ and $\{E_\al\}$ with a normalization $\ex{H_i,H_j}=\del_{ij}$ and $\ex{E_\al,E_\be}=\del_{\al,-\be}$. We can it to expand $\psi$ as
$$\psi=\sum_{i=1}^{r_{G}} \psi^i H_i + \sum_{\al\in  R_{\frak{g}_{\mathbb C};\frak{h}^c_{\mathbb C}} } \psi^\al E_\al.$$
Therefore, canonical quantization gives the commutation relations
$$[\hat \psi^{i} ,\hat \psi^{j}]=\del^{ij}\quad\text{and}\quad  [\hat \psi^\al , \hat \psi^{-\al}]=1,$$
To make sense of the quantization, we need to impose the following conditions as in Section \ref{sec:non-compact}
$$
(\hat \psi^i)^\dagger=\hat \psi^i,\quad   (\hat \psi^\al)^\dagger = \hat \psi^{-\al}.
$$
On the path integral side, this means that $\psi$ belongs to an appropriate middle dimensional contour of $ \Pi L {\mathfrak g_{\mathbb C}}$.

As in Section \ref{W-defects}, in the Hamiltonian representation the fermion path integral becomes the trace
\beq\label{int=trace-2}
 \bm\int \chi(A)e^{-\frac{1}{2}\int_{0}^{T}\ex{\psi,\pa_t^{\bm h/T}\psi} dt}\cD\psi= \text{Tr}_{\cH_{F,\mathfrak g}}\left[(-1)^F \hat\chi(A)\, e^{-i\hat H_{F}T}\right], 
 \eeq
and
\beq\label{H-F-3}
\begin{gathered}
\hat H_F=\sum_{\al \in  R^+_{\frak{g}_{\mathbb C};\frak{h}^c_{\mathbb C}}} \frac{i\ex{\al,\bm h}}{T} {\hat \psi^{\al} \hat \psi^{-\al}- \hat \psi^{-\al} \hat \psi^{\al}\ov 2}\\
=\sum_{\al \in  R^+_{\frak{g}_{\mathbb C};\frak{h}^c_{\mathbb C}}}  \frac{i\ex{\al,\bm h}}{T} \left(\hat \psi^{\al} \hat \psi^{-\al}- \frac{1}{2}\right).
\end{gathered}
\eeq
Now in this gauge, we have an identification $\hat \chi^i=\hat \psi^i$ with $i=1,\dots,r_G$ and $c_G=\pm i^{r_G(r_G-1)/2}$, and this makes the product of the inserted operator $\hat\psi(A)$ with $(-1)^{F}$ effectively as a projection operator to the reduced fermionic Hilbert space 
$$\cH_{\frak{g}/\frak{h}^{c}}=\bigotimes _{\al \in  R^+_{\frak{g}_{\mathbb C};\frak{h}^c_{\mathbb C}}} \CC^{2},$$
where each $\mathbb C^2$ is an irreducible module for the pair $\hat \psi_\al, \hat \psi_{-\al}$.  

For each factor $\mathbb C^2$ associated with a positive root $\al$, we  define the fermion number operator  by 
\beq \label{eq:NF}
\hat N_{F,\al}=  \hat \psi^{\al} \hat  \psi^{-\al}, \quad \al>0.
\eeq
Together with $  (\hat \psi^\al)^\dagger = \hat \psi^{-\al}$, this means that $\hat \psi^{\al}$ is a creation operator and $\hat \psi^{-\al}$ as an annihilation operator. As a consequence, 
$$-i\hat{H}_{F}T=\sum_{\al \in  R^+_{\frak{g}_{\mathbb C};\frak{h}^c_{\mathbb C}}} \ex{\al,\bm h} \left(\hat N_{F,\al}- \frac{1}{2}\right)$$
and 
$$\text{Tr}_{\cH_{F,\mathfrak g/\frak{h}^{c}} }\left[e^{-i\hat H_{F}T}\right]=\prod_{\al \in  R^+_{\frak{g}_{\mathbb C};\frak{h}^c_{\mathbb C}}}  \left(e^{- {\ex{\al, \bm h}\ov 2}} -e^{ {\ex{\al, \bm h}\ov 2}} \right).$$

Therefore, the fermionic path integral \eqref{eq:ferintgen} for $A=\bm{h}/T$ in the Hamiltonian formulation is unambiguously given by\footnote{One can also compute the fermionic path integral directly using the zeta function regularization analogous to \cite{Choi:2021yuz}. We note that for the consistent result, it is important to determine the path integral measure carefully, especially such that it takes into account our choice of the fermion number in \eqref{eq:NF}. }  
\beq \label{eq:deftau}
\tau\left( \bm h \right)\eqdef\prod_{\al \in  R^+_{\frak{g}_{\mathbb C};\frak{h}^c_{\mathbb C}}}  \left(e^{- {\ex{\al, \bm h}\ov 2}} -e^{ {\ex{\al, \bm h}\ov 2}} \right).
\eeq
The product can be written as a Pfaffian of the  following skew-symmetric linear operator acting on $\mathfrak g/\mathfrak h^c$
\beq
\tau\left( \bm h \right)=\left.\text{Pf}  \left( i (e^{-{1\ov 2} \text{ad}_{\bm h}} -e^{{1\ov 2} \text{ad}_{\bm h}} )\right)\right |_{\mathfrak g/\mathfrak h^c},
\eeq
where the sign convention of the Pfaffian is chosen to be  consistent with the definition  \eqref{eq:NF}  of the fermion number and is given by the right hand side of \eqref{eq:deftau}. Since $e^{\text{ad}_X}=\text{Ad}_{e^X}$, we can rewrite $\tau\left( \bm h \right)$ as
\beq
\tau\left( \bm h \right)=\text{Pf} \left( i\left( \text{Ad}_{ e^{-{1\ov 2} {\bm h}}}-\text{Ad}_{ e^{{1\ov 2} {\bm h}}} \right) \right)  \Big |_{\mathfrak g/\mathfrak h^c}.
\eeq
Finally, the holonomies of gauge equivalent connections are related by the adjoint action of $K$, so for a connection $\nabla_A$ in the adjoint bundle $P\times_{\text{Ad}|_K}\, \mathfrak g$ which is gauge equivalent to $\bm h/T$ we have
\beq
\tau\left( \bm h \right)=\text{Pf}{\,'} \left( i\left(\text{Hol}^{-1/2}_{S^1_T}(\nabla_A)-\text{Hol}^{1/2}_{S^1_T}(\nabla_A)\right) \right) .
\eeq

This proves \eqref{eq:ferintgen} and together with the DeWitt term originated from the supersymmetry algebra, completes the proof of the main identity \eqref{Tr-general}.

\begin{remark} The definition \eqref{eq:NF} is consistent with the choice of fermion number operator in case $G=\mathrm{SL}(2,\mathbb R)$ in Sections \ref{gauge-sigma}--\ref{W-defects}. Namely, we have (see Appendix \ref{A})
$$H=\frac{1}{2\sqrt{2}i}T_{3},\; E_{\alpha}=\frac{1}{4}(T_{1}+iT_{2}),\;E_{-\alpha}=\frac{1}{4}(T_{1}-iT_{2}),$$
so 
$$\la H,H\ra=\la E_{\al}, E_{-\al}\ra=1\quad\text{and}\quad \alpha(H)=\frac{1}{\sqrt{2}},\quad\la\alpha,\bm{h}\ra=2ih.$$
Therefore, formula \eqref{H-F-3} gives Hamiltonian \eqref{H-F-0}.
\end{remark}

\enlargethispage{3em}
Now we are ready to obtain the path-integral version of the pre-trace formula from \eqref{Tr-general} by carefully taking care of the path integral measure. We gauge fix the right hand side  of \eqref{Tr-general} by imposing $\dot{A}=0$ so that $A=\bm h/T$ with $\bm h \in \mathfrak k$. 
Then the gauge fixed path integral in terms of holonomy $u=e^{\bm h}\in K$ can be written as
\beq
\begin{gathered}
Z(iT)=\Tr[e^{-iT \Delta/2}]\\
={e^{-i\ex{\rho,\rho}T\ov 2}\ov \text{vol}(K)} \int_K du \bm\int   {1\ov \tau(\bm h)} \psi_{\mathfrak h^c}e^{i S_f[\bm h](g,\psi)} \mathscr D g\mathscr D\psi,
\end{gathered}
\eeq
where we denote $ \psi_{\mathfrak h^c}=\psi(\bm h/T)$. Note that since two connections with same holonomy are gauge equivalent, the integral does not depend on the choice of $\bm{h}$ in $u=e^{\bm h}$.

The second residual gauge symmetry, which is present for non-abelian $K$, is given global gauge transformations  $A\rightarrow kAk^{-1}$ with $k\in K$. Therefore we can gauge fix such that $\bm h\in \mathfrak t $ and $\bm t=e^{\bm h}\in T_K$, where $T_K$ is the maximal torus of $K$. Using the Weyl integration formula we obtain after the gauge fixing
\beq
\begin{gathered}
Z(iT)=\Tr[e^{-iT \Delta/2}] \\={e^{-i\ex{\rho,\rho}T\ov 2}\ov |W_K|\text{vol}(T_K)} \int_{T_K} d\bm t |\del(\bm t)|^2 \bm\int   {1\ov \tau(\bm h)} \psi_{\mathfrak h^c}e^{i S_f[\bm h](g,\psi)} \mathscr D g\mathscr D\psi.
\end{gathered}
\eeq
Here $W_K$ is the Weyl group of $K$ and the correct radial measure can be read off from
\beq
\del(\bm t)=\det(1-\text{Ad}_{\bm t})_{\mathfrak k/\mathfrak t} \quad\text{and}\quad  |\del(\bm t)|^2= \prod_{\al\in R_{   \mathfrak k_{\mathbb C},\mathfrak t_{\mathbb C}} }  \left(e^{\ex{\al,\bm h}\ov 2}-e^{-{\ex{\al,\bm h}\ov 2}}\right)^2.
\eeq

Finally, as in Section \ref{pre-trace}, we obtain the path integral pre-trace formula
 \beq\label{G-pretrace}
~Z(iT)&=\sum_{[\gamma]}Z_{[\gamma]}(iT)
\\&=\sum_{[\gamma]}{\text{vol}(\Gamma_\gamma\bk G_\gamma)} \int_{G_\gamma\bk G}  \mathcal F_\gamma (g_0;\lambda) dg_0,
\eeq
where
\beq
\begin{gathered}
\mathcal F_\gamma (g_0;\lambda)\\={e^{-{i \ex{\rho,\rho}T \ov 2}} \ov |W_K| \text{vol}(T_K)} \int _{T_K} {d\bm t} \, {|\del (\bm t)|^2 } \bm\int \limits _{\mathcal P_\gamma(G;g_0)\times \Pi L\frak{g} }{1\ov \tau(\bm h)} \psi_{\mathfrak h^c} e^{i S_f[\bm h]}\mathscr Dg\mathscr  D\psi \,.
\end{gathered}
\eeq

As in case of $G=\mathrm{SL}(2,\RR)$, we  compute each contribution to the pre-trace formula \eqref{G-pretrace} separately by using a supersymmetric localization with a $\gamma$-dependent deformation $S_f[\bm h]\rightarrow S_f[\bm h]+\lambda \del V_\gamma$.  As before, we see that conjugacy classes $[\gamma]$ and $[\gamma']$ give the same path integral contribution if $\g'=z\g$ where $z\in \mathcal Z(G)$, the center of $G$. This is clear geometrically since $K$  contains  $\mathcal{Z}(G)$.  Therefore, we divide the conjugacy classes in $\Ga$ by their action on $G/K$.

%\medskip
%\noindent
$\bullet$ {\it Conjugacy classes with trivial action}
%\medskip

They are of the form  $[\gamma]=z$, where $z\in \mathcal Z_\Gamma (G) =\mathcal Z(G) \cap \Gamma$, so we can choose $[\gamma]=I$. In this case the deformation of the same form as in Section~\ref{sbc: STF},
\beq
~& V_0=-\int dt {1\ov 2}\ex{\pa_t^{A_0} J,\pa_t^{A_0} \psi},
\\& \del V_0=\int dt {1\ov 2}\ex{\pa_t^{A_0} J,\pa_t^{A_0} J}+{i\ov 2}\ex{\pa_t^{A_0}\psi,\pa_t^J \pa_t^{A_0}\psi},
\eeq
where $A_0$ is regular element in $\mathfrak t$.
The isolated critical points are given by
\beq
g_{\{n_j \}}(t)=\exp \left({  {{n_j}  \sr_j t\ov T} }\right)g_0,
\eeq
where $\{ \sr_j\}$ is a basis of characteristic lattice $\Ga_{K}$ of the maximal torus, $T_K=\mathfrak t/\Gamma_{K}$. The computation is analogous to $G=\mathrm{SL}(2,\RR)$ case in Section \ref{sbc: STF}, and we get
\begin{equation} \label{eq:gentriv}
\begin{gathered}
Z_{I}(iT)={\text{vol}(\Gamma\bk G)e^{-{i \ex{\rho,\rho}T \ov 2}}\ov (2\pi i T)^{d_{\mathfrak p}/2} (-2\pi i T)^{d_{\mathfrak k}/2}  i^{ \text{dim}(R_{\frak{g}_{\mathbb C};\frak{h}_{\mathbb C}^c}^+) } |W_K| \text{vol}(T_K)}\\\times \int_{T_K}  { |\del(\bm t)|^2\ov \tau(\bm h)} \sum_{\sr \in \Ga_{K} } \left(\prod_{\al\in R_{\frak{g}_{\mathbb C};\frak{h}_{\mathbb C}^c}^+} -i \ex{\bm h-\sr,\al }\right)e^{{i\ex{\bm h-\sr ,\bm h-\sr }\ov 2T}}d\bm t,
\end{gathered}
\end{equation}      
where $\text{dim}(R_{\frak{g}_{\mathbb C};\frak{h}_{\mathbb C}^c}^+ )$ is a number of positive roots in $R_{\frak{g}_{\mathbb C};\frak{h}_{\mathbb C}^c}$.

Thus the total contribution of the conjugacy classes with trivial action on $G/K$ is $|\mathcal Z_\Gamma (G)| Z_{I}(iT)$.

\medskip
\noindent
$\bullet$ {\it Conjugacy classes with non-trivial action}
\medskip

For a representative $\g=e^{\bm r}$ of an element in $\Gamma\bk \mathcal Z_\Gamma (G)$  we choose the deformation
\beq 
~& V=-\int dt {1\ov 2}\ex{\pa^{A_\gamma}_t J,\pa_t^{A_\gamma} \psi},
\\& \del V=\int dt {1\ov 2}\ex{\pa_t^{A_\gamma} J,\pa_t^{A_\gamma} J}+{i\ov 2}\ex{\pa_t^{A_\gamma} \psi,\pa_t^J \pa_t^{A_\gamma}\psi},
\eeq
where $A_\gamma$ is a regular element of $\mathfrak t^\gamma\equiv \mathfrak h_\gamma^c  \cap \mathfrak k$ where $\mathfrak h^c_\gamma$ is the maximally compact Cartan subalgebra of the centralizer group $Z_{\mathfrak k}(\bm r) \subset G$. In this case, we choose $\mathfrak t$ and $\mathfrak h^c$ such that $\mathfrak h_\gamma^c \subset \mathfrak h^c$.

Note that we additionally introduced $A_\gamma$ in the deformation to ensure that the path integral localizes onto isolated critical points. Explicitly, the points in the localization locus are given by
\beq
g_{\gamma,\{n_j\}}(t)=\exp \left( {{(r+n_j  \sr^\gamma_j)t\ov T}} \right) g _0,\quad
\eeq
where $ \sr^\gamma_j$ is a basis of $\Ga_{K}^\gamma \equiv \Ga_{K} \cap \mathfrak t^\gamma$.  Thus for each non-trivial conjugacy class $[\gamma]$
we have the following contribution to the pre-trace formula \eqref{G-pretrace},
\begin{equation} \label{eq:gennontriv}
\begin{gathered}
Z_{[\gamma]}(iT)=\\{\text{vol}(\Gamma_\gamma\bk G_\gamma) \ov (2\pi i T)^{d_{\mathfrak p}/2} (-2\pi i T)^{d_{\mathfrak k}/2}   |W_K|\text{vol}(T_K)} \int_{T_K}   \left( \prod_{\al\in R^{+}_{\frak{g}_{\mathbb C};\frak{h}_{\mathbb C}^c}}  -i \ex{\bm h,\al} \right) {|\del(\bm t)|^2\ov \tau(\bm h)}d\bm t 
\\\times  \sum_{\sr^\gamma\in\Ga_{K}^\gamma} e^{{i(\ex{\bm h,\bm h}+\ex{\bm r+\sr^\gamma, \bm r+\sr^\gamma})\ov 2T}}
\left(\prod_{\al\in R^{+}_{\frak{g}_{\mathbb C};\frak{h}^{c}_{\mathbb C}}}{\ex{\bm r+\sr^\gamma,\al}\ov 2  i \sinh{\ex{\bm r+\sr^\gamma,\al}\ov 2}} \right) \int_{G_\gamma\bk G} e^{-{i\ex{\text{Ad}_g \bm h,\bm r+\sr^\gamma}\ov T}}dg.
\end{gathered}
\end{equation} 

As in Section \ref{sbc: STF}, we formally interchange the integration over $h$ and the summation over the set of critical points, and bypass the singularities in the integration over $h$ by choosing Feynman like  $r_{G}$-dimensional contour, which is consistent with the Weyl's law. 

\subsection{ Selberg trace formula for compact hyperbolic 3-manifold}

Here we consider another rank $1$ example  when $X=\Gamma\bk G/K$ is a compact hyperbolic 3-manifold. It corresponds to the case  $G=\mathrm{SL}(2,\mathbb C)$ and $K=\mathrm{SU}(2)$, so
$G/K=\HH^{3}$ is the Lobachevsky space, 
and $\Ga$ is a cocompact purely loxodromic Kleinian group. To simplify notations, here we consider the case  $-I\in \Gamma$ and hence $\mathcal Z_\Gamma (G)=\{I,-I\}$. Each element in $\Gamma$ that does not belong to $\mathcal Z_\Gamma (G)$ is loxodromic, and by definition it is conjugate to a unique element in $SL(2,\mathbb C)$ of the form
\beq \label{eq:loxo}
\begin{pmatrix} a_\gamma &0 \\ 0& a_\gamma^{-1} \end{pmatrix},\quad |a_{\g}|>1,
\eeq
and we denote $a_{\gamma}=e^{r+ib}$ where $r>0$ and $b\in [0,2\pi)$. Since $\mathcal Z_\Gamma(G)=\{I,-I\}$, the elements with $(r,b)$ and $(r,b+\pi)$ act identically on $\HH^{3}$. 

Firstly, we apply \eqref{eq:gentriv} to compute the contribution from the trivial elements which in our case $[\gamma]=I$ or $[\gamma]=-I$. We choose to parametrize \linebreak{$T_K\simeq U(1)$ by $\bm t=e^{\bm h}$} with $\bm h=h \cdot i\sigma_3$ where $h\in [0,2\pi)$. Regarding $\mathfrak g\simeq \mathfrak s\mathfrak l(2,\mathbb C) \simeq \mathfrak s \mathfrak o(3,1)$ as a real semisimple Lie algebra, we have $\mathfrak g_{\mathbb C}\simeq  \mathfrak s \mathfrak o(3,1)_{\mathbb C}\simeq \mathfrak s \mathfrak o(4,\mathbb C) $, and an elementary computation gives
\beq
|\del(\bm t)|^2=4\sin^2h \quad\text{and}\quad \tau(\bm h)=-4 \sin^2 h.
\eeq
As in Section \ref{sbc: STF}, we can combine the sum over $\Gamma_K$ and the integral over $T_K$ to a single integral over $\mathbb R^{r_K}=\mathbb R$, and we obtain
\beq
Z_{I}(iT)={\text{vol}(\Gamma\bk G)e^{-{i T \ov 8}}\ov (2\pi T)^{3} \pi } \int_{\mathbb R} dh \,h^2  e^{-{4ih^2\ov T}}.
\eeq
Analytic continuation and evaluation of  the resulting integral gives 
\beq
Z_{I}(\be)&= \text{vol}(\Gamma\bk G)e^{-{\be \ov 8}}{ 1\ov 128\pi^{7/2} \be^{3/2} },
\eeq 
 where we emphasize that the volume is in terms of the Riemannian measure of the Cartan-Killing metric on $G$ defined in \eqref{eq:CKmetric}. Since $\Gamma \cap K=\mathcal Z_{\Gamma}(G)=\{I,-I\}$ the relation between volumes \eqref{eq:volumerelation} still holds in this case, and using $\text{vol}(K)=32\sqrt{2}\pi^2 $, we have
 \beq
Z_{I}(\be)={\text{vol}(X)e^{-{\be \ov 8}}\ov 2 (2\pi \be)^{3/2}}
 \eeq
 
Secondly, we apply \eqref{eq:gennontriv} to compute the contribution from the loxodromic elements. Each $\gamma\in\Gamma$ in a form \eqref{eq:loxo} is conjugated to $e^{\bm r}$, where $\bm r=(r+ib) \sigma_3$, and the critical points are given by 
\beq
g_{\gamma,n}=\begin{pmatrix}  e^{{r+i(b+2\pi n)\ov T}} &0\\0& e^{-{r+i(b+2\pi n)\ov T}} \end{pmatrix},\quad n\in\ZZ.
\eeq
In other words, $\Gamma_K^\gamma\simeq \mathbb Z$ and $\sr^r=2\pi n i \sigma_3$, where $n\in \mathbb Z$. Therefore, formula~\eqref{eq:gennontriv} becomes
\beq \label{eq:H3e1}
\begin{gathered}
Z_{[\gamma]}(iT) \\={\text{vol}(\Gamma_\gamma \bk G_\gamma )e^{-{i T \ov 8}}\ov 8 \pi^4 T^{3}} \int_{-\pi}^\pi dh \, \sum_{n\in\mathbb Z}
 e^{{ 8i (r^2 -(b+2\pi n)^2 -h^2)\ov T} } h^2  {(r^2+(b+2\pi n))^2\ov \vert \sinh^2{ (r+ib)}\vert^2}
 \\ \times  \int_{G_\gamma\bk G} e^{-{i\ex{\text{Ad}_g \bm h,\bm r+\sr^\gamma}\ov T}}dg.
 \end{gathered}
\eeq

To compute the orbital integral, we use the Iwasawa decomposition $G=ANK$ in the form
\beq
g&=e^{ a T_1} \exp{ \begin{pmatrix} 1& (n_r+i n_i)  \\ 0&1\end{pmatrix}} e^{i \phi \sigma_3}e^{i\theta\sigma_2}e^{i\psi\sigma_3}
\eeq
where $\phi\in[0,\pi),~ \theta\in[0,\pi/2),~\psi\in[0,2\pi)$ are the standard Euler angles for $\mathrm{SU}(2)$. The corresponding Haar measure on $G$ with respect to the Cartan-Killing metric (see Remark \ref{measure}) is 
   \beq
 dg= c_G da dn_r dn_i d\phi d\theta d\psi ,\quad c_G= 2^{10}\sin(2\theta).
   \eeq
To determine the measure of $G_\gamma\bk G$ where $G_\gamma=\{e^{(a+ib)\sigma_3}| \,a\in \mathbb R,\,b\in [0,2\pi) \}$, we use the property of the integrand in the orbital integral is invariant under the transformation $g\rightarrow f g$ for any $f\in G_\gamma$. As in Section \ref{sbc: STF}, we have a freedom to choose an overall normalization of the Haar measure on $G_\gamma$ which will not affect the final result. Here we choose $df={1\ov 2\pi} da db$, and therefore the measure of the orbital integral over $G_\gamma\bk G$ effectively becomes $c_G dn_r dn_i d\phi d\theta d\psi$.

The integration over $n_r$ and $n_i$ gives delta functions and the orbital integral becomes
\beq
\begin{gathered}
\int_{G_\gamma\bk G} e^{-{i\ex{\text{Ad}_g \bm h,\bm r+\sr^\gamma}\ov T}}dg \\=   {(2\pi )^3 T^2\ov  h^2(r^2+(b+2\pi n)^2)} (e^{i16(b+2\pi n)h} +e^{-i16(b+2\pi n)h}).
\end{gathered}
\eeq
Interchanging the sum over $n$ and the integral over $h$ in \eqref{eq:H3e1}, we get
\beq \label{eq:H3e2}
\begin{gathered}
 Z_{[\gamma]}(iT)\\={\text{vol}(\Gamma_\gamma \bk G_\gamma )e^{-{i T \ov 8}}\ov \pi T} \int_{\mathbb R} dh \, ( 
 e^{{ 8i (r^2 -(b+h)^2)\ov T} }+e^{{ 8i (r^2 -(b-h)^2)\ov T} }){1\ov \vert \sinh{ (r+ib)}\vert^2} 
 \\={r_0e^{-{iT \ov 8}}  } {1 \ov2 \sqrt { 2\pi i T} }e^{ {8 i r^2 \ov T}} {1 \ov \vert \sinh{ (r+ib)}\vert^2},
 \end{gathered} 
\eeq
where we used $\text{vol}(\Gamma_\gamma \bk G_\gamma )=r_0/2$ where $r_{0}$ corresponds to a primitive element $\gamma_0$ of $\gamma$ with $a_{\gamma_0}=e^{r_0+ib_0}$.

Finally, upon the analytic continuation we get
\beq
 Z_{[\gamma]}(\be)= {1\ov 2\sqrt{ 2\pi \be} }{r_0 \ov \vert \sinh{ (r+ib)}\vert^2} e^{ -{8 r^2 \ov \be}-{\be \ov 8}}
\eeq

Therefore, we obtain the Selberg trace formula on a compact hyperbolic 3-manifold, 
\beq
\Tr[e^{-{\be \Delta\ov 2 }}]&={\text{vol}(X)e^{-{\be \ov 8}}\ov (2\pi \be)^{3/2}} +\sum_{\substack{ [\gamma]\in\text{loxodromic} \\ 0\leq \text{arg}(a_\gamma)<\pi }} {1\ov \sqrt{ 2\pi \be} }{r_0 \ov \vert \sinh{ (r+ib)}\vert^2} e^{ -{8 r^2 \ov \be}-{\be \ov 8}},
\eeq
where for loxodromic elements we simplified the domain to be $0\leq \text{arg}(a_\gamma)<\pi$ using the property $Z_{[\gamma]}=Z_{[-\gamma]}$. After rescaling the Riemannian metric on $\HH^{3}$ to have the scalar curvature $-1$, we obtain the Selberg trace formula in the form used in the literature \cite{elstrodt2013groups} (see also \cite{friedman2005selberg}).

\begin{appendix}

\section{Laplace operators on $\mathrm{SL}(2,\RR)$ and $\mathrm{SL}(2,\RR)/\mathrm{SO}(2)$}\label{A}
Let $T_1=\sigma_3$, $T_2=\sigma_1$, $T_3=i\sigma_2$ be the basis of the Lie algebra $\mathfrak{g}=\mathrm{sl}(2,\RR)$, so that
$\frak{p}$ is generated by $T_{1}, T_{2}$ and $\frak{k}$ --- by $T_{3}$. In this basis the Killing metric $$g_{ab}=\ex{T_{a},T_{b}}=4\Tr T_{a}T_{b}$$  
is diagonal and
\beq\label{killing}
g_{11}=g_{22}=8,\;g_{33}=-8.
\eeq

The Killing form on $\mathfrak g$ naturally induces the Cartan-Killing metric on $G$ as $ds^2=4\Tr(g^{-1}dg \, g^{-1}dg)$.
One natural coordinates $(x,y,\theta)$ on \linebreak{$G=\mathrm{SL}(2,\RR)=\RR\times\RR_{>0}\times S^{1}$} are given by the Iwasawa decomposition $G=ANK$, 
$$g=\begin{pmatrix}\sqrt{y} & 0\\
0 & \frac{1}{\sqrt{y}}\end{pmatrix}\begin{pmatrix}1 & \frac{x}{y}\\
0 & 1\end{pmatrix}\begin{pmatrix}\;\;\cos\theta & \sin\theta\\
\!-\sin\theta &\cos\theta\end{pmatrix}.$$
In the path integral computations in the main text, we need to use a volume form $dg$ for the pseudo-Riemannian metric on $G$, defined by the Cartan-Killing form. Using the Iwasawa decomposition and \eqref{killing}, we obtain
\begin{equation}\label{Haar}
dg=4\sqrt{2}\,\frac{dxdyd\theta}{y^{2}}.
\end{equation}
 Correspondingly, the Riemannian volume form on $K<G$ is $dk=2\sqrt{2}d\theta$, so that the induced volume form on $\HH=G/K$ is twice the hyperbolic area form 
$$d\mu=\frac{dxdy}{y^{2}}.$$
In particular, for $X=\Gamma\bk G/K$ we have $\mu(X)=\text{vol}(X)/2$.

Defining left-invariant vector fields on $G$ by 
$$ (\hat{X}f)(g)=\left.\frac{d}{dt}\right|_{t=0}f(ge^{tX}),\quad X\in\frak{g},$$
we have
\begin{align*}
\hat{T}_{1} & = -2y\sin 2\theta\frac{\pa}{\pa x}+2y\cos 2\theta\frac{\pa}{\pa y} + \sin 2\theta\frac{\pa}{\pa\theta},\\
\hat{T}_{2} & = 2y\cos 2\theta\frac{\pa}{\pa x}+2y\sin 2\theta\frac{\pa}{\pa y} - \cos 2\theta\frac{\pa}{\pa\theta},\\
\hat{T}_{3} &=\frac{\pa}{\pa\theta},
\end{align*}
so
$$\Delta=-\frac{1}{8}(\hat{T}^{2}_{1}+\hat{T}^{2}_{2}-\hat{T}^{2}_{3})=-\frac{y^{2}}{2}\left(\frac{\pa^{2}}{\pa x^{2}}+\frac{\pa^{2}}{\pa y^{2}}\right) + \frac{y}{2}\frac{\pa^{2}}{\pa x\pa\theta}.$$
Note that operator $\Delta$ is not elliptic.

Restriction of $2\Delta$ to the subspace of functions $f$ on $G$ satisfying 
$$f(ge^{\theta T_{3}})=e^{in\theta}f(g),\quad n=2k\in\ZZ,$$ is the so-called Maass Laplacian of weight $k$\footnote{We parameterized Laplacians by $k\in\frac{1}{2}\ZZ$ instead of $n=2k\in\ZZ$ since they effectively act on $k$-differentials on $X$.} 
, elliptic operator $D_{k}/2$, where 
 $$D_{k}f =-y^{2}\left(\frac{\partial^{2}f}{\partial x^{2}}+\frac{\partial^{2}f}{\partial y^{2}}\right)+2iky\frac{\partial f}{\partial x}.$$

Operator $D_{k}$ acts the Hilbert space $L^{2}_{k}(X)$ of functions $f(z)$ on $\mathbb{H}$, satisfying
$$
f\left(\frac{az+b}{cz+d}\right)\frac{|cz+d|^{2k}}{(cz+d)^{2k}}=f(z),\quad \g=\begin{pmatrix} a &b\\ c&d\end{pmatrix}\in\Gamma
$$
and square integrable over the fundamental domain for $\Gamma$ with respect to hyperbolic area element $y^{-2}dxdy$.
When $k$ is half-integer ($n=2k$ is odd), then we assume $\Gamma$ does not contain $-I$\footnote{Note that in $X=\Gamma\bk\HH$ the Fuchsian group $\Gamma$ is a subgroup of $\mathrm{PSL}(2,\RR)$, and there $2^{2g}$ its lifts to $\mathrm{SL}(2,\RR)$ that do not contain $-I$, provided all elliptic generators are of odd order.}, or twist with a one-dimesnional representation $\pi$ of $\Ga$ such that $\pi(-I)=-1$.

One also considers Hodge Laplacian, the $\bar\partial$-Laplace operator 
$$\Delta_{k}=-4 y^{2-2k}\frac{\partial}{\partial z}y^{2k}\frac{\partial}{\partial \z},$$
acting on the Hilbert space $\curly H_{2k}(\Gamma)$ of automorphic forms of weight $2k$ for group $\Gamma$, functions $f(z)$ satisfying
\begin{equation}\label{n-0}
f\left(\frac{az+b}{cz+d}\right)=(cz+d)^{2k}f(z),\quad \g=\begin{pmatrix} a &b\\ c&d\end{pmatrix}\in\Gamma
\end{equation}
and square-integrable with respect to the measure $y^{2k-2}dxdy$. When $n$ is odd, the same conditions on $\Gamma$ are imposed. Functions $f\in\curly H_{2k}(\Gamma)$ corresponds to $k$-differentials $f(z)dz^{k}$ on a Riemann surface $X=\Gamma\bk\HH$, spinors for $k=1/2$.

The operators $\Delta_{k}$ are non-negative and the isometry 
$$f(z)\mapsto y^{k}f(z)$$
between Hilbert spaces $\curly H_{2k}(\Gamma)$ and $L^{2}_{k}(X)$ conjugates $\Delta_{k}$ and $D_{k} +k(k-1)$, so they have the same spectrum. 

Introducing Maass operators
$$K_{k}=(z-\z)\frac{\partial}{\partial z} +k,\quad L_{k}=-(z-\z)\frac{\partial}{\partial \z}-k,$$
we have
$$D_{k+1}K_{k}=K_{k}D_{k},\quad D_{k-1}L_{k}=L_{k}D_{k}.$$
From these equations it follows that the spectra of operators $D_{k}$ and $D_{k+1}$ coincide, except for possible finitely many eigenvalues related to zero modes of Hodge Laplacians (see \cite{fay1977fourier}).

\end{appendix}

\bibliographystyle{amsplain}
\bibliography{Ref.bib}

\end{document}